\documentclass[format=sigconf]{acmart}

\usepackage[utf8]{inputenc}
\usepackage{natbib}
\usepackage{amsmath}
\usepackage{amsfonts}
\usepackage{xcolor}
\usepackage{graphicx}
\usepackage{wrapfig}
\usepackage{cleveref}

\usepackage{listings}

\def\withcolor{}

\ifdefined\withcolor
	\definecolor{haskellblue}{rgb}{0.0, 0.0, 1.0}
	\definecolor{haskellred}{rgb}{0.9, 0.4, 0.1}
	\definecolor{gray_ulisses}{gray}{0.55}
	\definecolor{castanho_ulisses}{rgb}{0.0,0.4,0.0}
	\definecolor{preto_ulisses}{rgb}{0.41,0.20,0.04}
	\definecolor{green_ulisses}{rgb}{0.8,0.0,0.8}
    \definecolor{charcoal}{rgb}{0.21, 0.27, 0.31} 
\else
	\definecolor{haskellblue}{gray}{0.1}
	\definecolor{haskellred}{gray}{0.1}
	\definecolor{gray_ulisses}{gray}{0.1}
	\definecolor{castanho_ulisses}{gray}{0.1}
	\definecolor{preto_ulisses}{gray}{0.1}
	\definecolor{green_ulisses}{gray}{0.1}
	\definecolor{charcoal}{gray}{0.1}
\fi

\def\incodesize{\small}
\def\codesize{\small}

\lstdefinelanguage{HaskellUlisses} {
    basicstyle=\ttfamily\codesize\color{charcoal},
	sensitive=true,
	mathescape=true,
	morecomment=[l][\color{gray_ulisses}\ttfamily\codesize]{--},
	morestring=[b]",
	literate={ 
		{dollar}{{$\$$}}1
		{->}{{$\rightarrow$}}2
		{<-}{{$\leftarrow$}}2
		}, 
	stringstyle=\color{haskellred},
	showstringspaces=false,
	numberstyle=\codesize,
	numberblanklines=true,
	showspaces=false,
	breaklines=true,
	showtabs=false,
	emph=
	{[1]
		FilePath,IOError,acos,acosh,all,any,appendFile,approxRational,asTypeOf,asin,
        vcEmpty,vcTick,vcCombine,vcLessEqual,vcLess,
        mVC,mSender,mRaw,histVC,finAsc,
        pVC,pID,pDQ,pHist,
        zipWith,max,and,zipWith3,elem,flip,foldr,swap,
        newIORef,modifyIORef,atomicModifyIORef,
        receive,deliver,broadcast,
        pEmpty,dequeue,deliverable,deliverableHelper,
        proofConst,broadcastAlwaysDelivers,
        tailForHead,processOrder,
        step,xStep,
        stepLCD,
        xStepCD,
        receiveLCDpres,
        deliverLCDpres,
        broadcastLCDpres,
        len, vcCombineComm,
        happensBefore
	},
	emphstyle={[1]\color{haskellblue}},
	emph=
	{[2]
		bool,char,int,nat,int8,list,olist,incList,decList,pair,incPair,Bool,Char,Double,Either,Float,IO,Integer,Int,Maybe,Ordering,Rational,Ratio,ReadS,ShowS,String,
		Word8,Nat,NonZero,Nat64,Text,ByteString,ByteStringSZ,ByteStringN,
    Ptr,ForeignPtr,CSize,
    InPacket,Tree,Prop,TreeEq,TreeLt,Vec,
    NullTerm,IncrList,DecrList,UniqList,BST,MinHeap,MaxHeap,
    PtrN,ByteStringN,ByteStringEq,VO,ByteStringsEq,ByteStringNE,
		List,Even,
    PID,VC,M,Event,H,P,Op,
    ProcessLocalCausalDelivery,PLCD,
    OddInt,EvenInt,Proof,
    KvCommand,Key,Val,Value,NodeState,KvState,Map,
    Process,Event,Message,Comm
	},
	emphstyle={[2]\color{castanho_ulisses}},
	emph=
	{[3]
		case,class,data,deriving,do,else,if,switch,return,import,in,infixl,infixr,instance,val,let,rec,
		requires,ensures,assume,val,def,
		module,measure,predicate,of,primitive,then,refinement,type,where,lazy,do, import
	},
	emphstyle={[3]\color{preto_ulisses}\textbf},
	emph=
	{[4]
		quot,rem,div,mod,notElem,seq,
	},
	emphstyle={[4]\color{castanho_ulisses}\textbf},
	emph=
	{[5]
		PS,Tip,Node,EQ,true,false,GT,Just,LT,Left,Nothing,Right,Show,Eq,Ord,Num,
		Cons,Nil,OCons,ONil,
        Broadcast,Deliver,
        OpReceive,OpBroadcast,OpDeliver,
        KvPut,KvDel,
	},
	emphstyle={[5]\color{green_ulisses}}
}

\lstnewenvironment{fscode}
{\lstset{language=HaskellUlisses,basicstyle=\ttfamily\small}}
{}

\lstnewenvironment{minted}
{\lstset{language=HaskellUlisses}}
{}

\lstnewenvironment{codebox}
{\lstset{language=HaskellUlisses,frame=tlbr,basicstyle=\ttfamily\codesize}}
{}

\lstMakeShortInline[language=HaskellUlisses,keepspaces=true,basicstyle=\ttfamily\incodesize]@

\newcommand{\hs}[1]{\lstinline[language=HaskellUlisses,keepspaces=true,basicstyle=\ttfamily\incodesize]{#1}}

\copyrightyear{2022}
\acmYear{2022}
\setcopyright{rightsretained}
\acmConference[IFL 2022]{Symposium on Implementation and Application of Functional Languages}{August 31-September 2, 2022}{Copenhagen, Denmark}
\acmBooktitle{Symposium on Implementation and Application of Functional Languages (IFL 2022), August 31-September 2, 2022, Copenhagen, Denmark}\acmDOI{10.1145/3587216.3587222}
\acmISBN{978-1-4503-9831-2/22/08}

\usepackage{booktabs}   
\usepackage{subcaption} 

\newtheorem{definition}{Definition}
\newtheorem{theorem}{Theorem}

\newcommand{\mi}[1]{\mathit{#1}}

\newcommand{\hbR}[2]{\ensuremath{#1 \rightarrow_{hb} #2}}
\newcommand{\poR}[3]{\ensuremath{#1 \rightarrow_{#3} #2}} 
\newcommand{\vcleR}[2]{\ensuremath{#1 \leq_{vc} #2}}
\newcommand{\vcltR}[2]{\ensuremath{#1 <_{vc} #2}}
\newcommand{\vcF}[1]{\ensuremath{\mi{VC}(#1)}}
\newcommand{\broadcastE}[1]{\ensuremath{\mi{broadcast}(#1)}}
\newcommand{\deliverE}[2]{\ensuremath{\mi{deliver}_{#1}(#2)}}

\usepackage{xspace}

\newcommand\hb{happens before\xspace}

\newcommand\hbr{happens-before relation\xspace}

\newcommand\po{process order\xspace}

\newcommand\por{\po relation\xspace}

\newcommand\vc{vector clock\xspace}

\newcommand\VC{Vector Clock\xspace}
\newcommand\vcp{\vc protocol\xspace}

\newcommand\VCP{\VC Protocol\xspace}

\newcommand\cd{causal delivery\xspace}
\newcommand\Cd{Causal delivery\xspace}
\newcommand\CD{Causal Delivery\xspace}
\newcommand\plcd{local \cd}
\newcommand\Plcd{Local \cd}
\newcommand\PLCD{Local \CD}

\newcommand\cba{causal broadcast protocol\xspace}

\newcommand\CBA{Causal Broadcast Protocol\xspace}

\begin{document}

\title{Verified Causal Broadcast with Liquid Haskell}
\author{Patrick Redmond}
\affiliation{\institution{University of California, Santa Cruz} \country{USA}}
\author{Gan Shen}
\affiliation{\institution{University of California, Santa Cruz} \country{USA}}
\author{Niki Vazou}
\affiliation{\institution{IMDEA} \country{Spain}}
\author{Lindsey Kuper}
\affiliation{\institution{University of California, Santa Cruz} \country{USA}}

\begin{abstract}
    Protocols to ensure that messages are delivered in \emph{causal order} are
    a ubiquitous building block of distributed systems.  For instance,
    distributed data storage systems can use causally ordered message delivery to ensure causal
    consistency,
    and CRDTs can rely on the existence of an
    underlying causally-ordered messaging layer
    to 
    simplify their implementation.
    A causal delivery protocol
    ensures that when a message is delivered to a process, any causally
    preceding messages sent to the same process have already been delivered to
    it.  While causal delivery protocols are widely used,
    verification of their correctness is less common,
    much less machine-checked proofs about executable implementations.

    We implemented a standard causal broadcast protocol in Haskell and used the
    Liquid Haskell solver-aided verification system to express and mechanically
    prove that messages will never be delivered to a process in an order that
    violates causality. We express this
    property using refinement types and prove that it holds of our
    implementation, taking advantage of Liquid Haskell's underlying
    SMT solver to automate parts of the proof and using its
    manual theorem-proving features for the rest.  We then put our
    verified causal broadcast
    implementation to work as the foundation of a distributed key-value store.
\end{abstract}

\maketitle

\section{Introduction}
\label{sec_introduction}

\emph{Causal message delivery}~\citep{birman-virtual-synchrony, schiper-causal-ordering, birman-reliable, birman-lightweight-cbcast} is a fundamental communication abstraction for distributed computations in which processes communicate by sending and receiving messages.
One of the challenges of implementing distributed systems is the asynchrony of message delivery; messages arriving at the recipient in an unexpected order can cause confusion and bugs.
A causal delivery protocol can ensure that,
when a message $m$ is delivered to a process $p$,
any message sent ``before'' $m$
(in the sense of \citeauthor{lamport-clocks}'s
``happens-before''; see \Cref{subsec_bg_system_model})
will have already been delivered to $p$.
When a mechanism for causal message delivery is available,
it simplifies the implementation of many important distributed algorithms,
such as replicated data stores that must maintain
causal consistency~\cite{ahamad-causal-memory, lloyd-cops},
conflict-free replicated data types~\cite{shapiro-crdts},
distributed snapshot protocols~\citep{acharya-causal-snapshots, alagar-causal-snapshots},
and applications that ``involve human interaction and consist of large numbers of communication endpoints''~\citep{van-renesse-controversy}.
A particularly useful special case of causal delivery is causal \emph{broadcast},
in which each message is sent to all processes in the system.
For example, a causal broadcast protocol enables a straightforward implementation strategy
for a causally consistent replicated data store --- one of the strongest
consistency models available for applications that must
maximize availability and tolerate network
partitions~\cite{mahajan-consistency}.
Conflict-free replicated data types (CRDTs) implemented in the \emph{operation-based}
style~\citep{shapiro-crdts, shapiro-crdts-comprehensive}
typically also assume the existence of an underlying causal broadcast layer~\citep[\S2.4]{shapiro-crdts}.

\begin{figure}
  \includegraphics[width=\columnwidth]{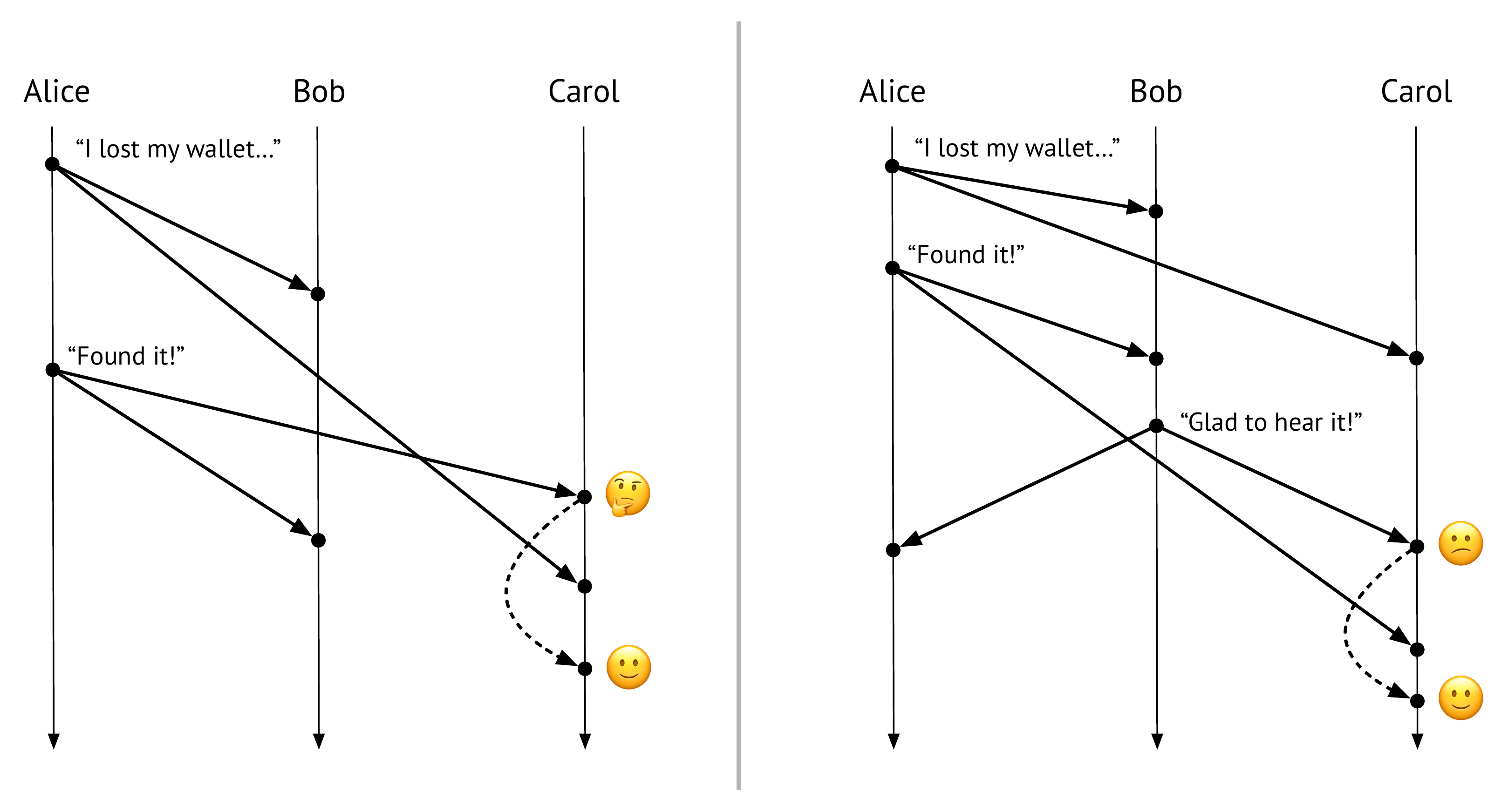}
  \caption{Two executions that violate \cd (\Cref{def_math_cd}).
    On the left, Carol sees Alice's messages in the
    opposite order of how they were sent.  On the right, Carol sees Bob's
    message before seeing Alice's second message.  The dashed arrows in both
    diagrams depict how a \cd mechanism (\Cref{subsec_bg_causal_broadcast})
    might delay received messages in a buffer for later delivery.}
  \label{fig:fifo-causal-violations}
\end{figure}

What can go wrong in the absence of causal broadcast?
Suppose Alice, Bob, and Carol are exchanging group text
messages.
Alice sends the message ``I lost my wallet...'' to the group,
then finds the missing wallet between her couch cushions
and follows up with a ``Found it!''
message to the group.  In this situation, depicted in Figure~\ref{fig:fifo-causal-violations}~(left), Alice has a reasonable expectation that Bob and Carol will see the messages in the order that she sent them, and such \emph{first-in first-out (FIFO) delivery} is an aspect of causal message ordering.  While FIFO delivery is already enforced\footnote{TCP's FIFO ordering guarantee applies so long as the messages in question are sent in the same TCP session.  Across sessions, additional mechanisms are necessary.} by standard networking protocols such as TCP~\citep{RFC0793}, it is not enough to eliminate all violations of causality.  In an execution such as that in Figure~\ref{fig:fifo-causal-violations}~(right), FIFO delivery is observed, and yet Carol sees Bob's message only after having seen Alice's initial ``I lost my wallet...'' message, so from Carol's perspective, Bob is being rude.  The issue is that Bob's ``Glad to hear it!'' response \emph{causally depends} on Alice's second message of ``Found it!'', yet Carol sees ``Glad to hear it!'' first.  What is called for is a mechanism that will ensure that, for every message that is applied at a process, all of the messages on which it causally depends --- comprising its \emph{causal history} --- are applied at that process first, regardless of who sent them.

One way to address the problem is to buffer messages at the receiving end until all causally preceding broadcast messages have been applied.  
The dashed arrows in \Cref{fig:fifo-causal-violations} represent the
behavior of such a buffering mechanism.
A typical implementation strategy is to have the sender of a message augment
the message with metadata (for instance, a \emph{\vc}; see
\Cref{subsubsec_bg_vc}) that summarizes that message's causal
history in a way that can be efficiently checked on the receiver's end to
determine whether the message needs to be buffered or can be applied
immediately to the receiver's state.
Although such mechanisms are well-known in the distributed systems
literature~\citep{birman-virtual-synchrony, birman-reliable,
birman-lightweight-cbcast}, their implementation is ``generally very delicate
and error prone''~\citep{bouajjani-verifying-cc}, motivating the need for
machine-verified implementations of causal delivery mechanisms that are usable
in real, running code.

To address this need, we use the Liquid Haskell~\citep{vazou-lh} platform to implement and verify the correctness of a well-known causal broadcast protocol~\citep{birman-lightweight-cbcast}.  Liquid Haskell is an extension to the Haskell programming language that adds support for \emph{refinement types}~\citep{rushby-predicate-subtyping, xi-array-dependent}, which let programmers specify logical predicates that restrict, or refine, the set of values described by a type.  Beyond giving more precise types to individual functions, Liquid Haskell's \emph{reflection}~\cite{vazou-refinement-reflection, vazou-theorem-proving-for-all} facility lets programmers use refinement types to extrinsically specify properties that can relate multiple functions~(see Section~\ref{subsec_lh}), and then prove those properties by writing Haskell programs to inhabit the specified types.  We use this capability to prove that in our causal broadcast implementation, processes deliver messages in causal order, ruling out the possibility of causality-violating executions like those in Figure~\ref{fig:fifo-causal-violations}.

We express causal delivery as a refinement type.  By doing so, we can take advantage of Liquid Haskell's underlying SMT automation where possible, while still availing ourselves of the full power of Liquid Haskell's theorem-proving capabilities via reflection where necessary.
A further advantage of Liquid Haskell as a verification platform is that it results in \emph{immediately executable} Haskell code, with no extraction step necessary, as with proof assistants such as Coq~\cite{bertot-coq} or Isabelle~\cite{wenzel-isabelle} --- making it easy to integrate our library with existing Haskell code.

Our causal broadcast implementation is a Haskell library that can be
used in a variety of applications.
While previous work has mechanically verified the correctness
of applications of causal ordering in distributed systems (such as
causally consistent distributed key-value stores~\cite{lesani-chapar, gondelman-causal-consistency-aneris}), factoring the causal broadcast protocol out into its own
standalone, verified component means that it can be reused in each of these
contexts.  There is a need for such a standalone component: for instance,
recent work on mechanized verification of CRDT
convergence~\cite{gomes-verifying-sec} \emph{assumes} the existence of a
correct causal broadcast mechanism for its convergence result to hold.
Our
separately-verified library could be plugged together with such verified CRDT implementations to get an end-to-end correctness
guarantee.
Therefore our library enables \emph{modular} verification of higher-level properties for applications built on top of the causal broadcast layer.
While recent work~\citep{nieto-crdts-aneris} takes precisely such a modular approach to verification of applications that use causal broadcast, our work is to the best of our knowledge the first to do so by expressing causal message delivery as a refinement type and leveraging SMT automation.

We make the following specific contributions:

\begin{itemize}

    \item
        We identify \emph{\plcd},
        a property that allows us to reduce the problem of determining that a distributed execution observes causal delivery to one that can be verified using information locally available at each process
        (\Cref{subsec_bg_verification_task}).

    \item
        We identify design choices that make a standard \cba amenable to verification.
        In particular,
        we implement the protocol in terms of a state transition system,
        and we implement message broadcast in terms of message delivery,
        leading to a simpler proof development (\Cref{subsec_impl_causal_broadcast}).

      \item
        We present novel encodings of \plcd and \cd as refinement types, and we give a mechanized proof that our causal broadcast library implementation satisfies the causal delivery property (\Cref{sec_verification}).

\end{itemize}
To evaluate the practical usability of our library, we put it to work as the foundation of a distributed in-memory key-value store and empirically evaluate its performance when deployed to a cluster of geo-distributed nodes (Section \ref{sec_demo}).
\Cref{sec_related_work} contextualizes our contributions with respect to existing
research, and \Cref{sec_conclusion} summarizes our work.  All of our code, including our causal broadcast library, our proof development, and our key-value store case study, is available at \url{https://github.com/lsd-ucsc/cbcast-lh}.

\section{System Model and Verification Task}
\label{sec_background}

In this section, we describe our system model (\Cref{subsec_bg_system_model}) and the \cba that we implemented and verified (\Cref{subsec_bg_causal_broadcast}), and we define the property that we need to show holds of our implementation (\Cref{subsec_bg_verification_task}).

\subsection{System Model}
\label{subsec_bg_system_model}
\label{def_math_process_order}

We model a distributed system as a finite set of $N$ \emph{processes} (or
\emph{nodes}) $p_i$, $i: 1..N$, distinguished by process identifier $i$.
Processes communicate with other processes by sending and receiving
\emph{messages}.
In our setting, all messages are \emph{broadcast} messages,
meaning that they are sent to all processes in the system,
including the sender itself.%
\footnote{For simplicity, we omit the messages that processes send to
themselves from examples in Figures~\ref{fig:fifo-causal-violations},
~\ref{fig:vc-definition}, and~\ref{fig:vector-clocks}.  We assume that these
self-sent messages are sent and delivered in one atomic step on the sender's
process.}
Our network model is \emph{asynchronous}, meaning that sent messages can take
arbitrarily long to be received.
Furthermore, for our safety result we need not assume that sent messages are eventually
received, so our network is also \emph{unreliable}
(although such an assumption would be necessary for liveness; see
Section~\ref{subsec_verif_liveness} for a discussion).

We distinguish between message receipt and message delivery:
processes can \emph{receive} messages at any time and in any order, and
they may further choose to \emph{deliver} a received message,
causing that message to take effect at the node receiving it
and be handed off to, for example, the user application running on that node.
Importantly,
although nodes cannot control the order in which they receive messages,
they can control the order in which they deliver those messages.
Imagine a ``mail clerk'' on each node that intercepts incoming messages and
chooses whether, and when, to deliver each one (by handing it off to the above
application layer and recording that it has been delivered).
We must ensure that the mail clerk delivers the messages in an
order consistent with causality, regardless
of the order in which messages were received --- implementing the behavior
illustrated by the dashed arrows in \Cref{fig:fifo-causal-violations}.

For our discussion of \cd, we need to consider two kinds of \emph{events} that occur on processes:
\emph{broadcast} events and \emph{deliver} events.
We will use \broadcastE{m} to denote an event that sends a message $m$ to all
processes,\footnote{Although a broadcast message has $N$ recipients, and may be
implemented as $N$ individual unicast messages under the hood, we treat the
sending of the message as a single event on the sender's process.}
and \deliverE{p}{m} to denote an event that delivers $m$ on process $p$.
We refer to the totally ordered sequence of events that have occurred on a process $p$ as
the \emph{process history}, denoted $h_p$.
For events $e$ and $e'$ in a process history $h_p$, $e$ and $e'$ are in \emph{\po}, written \poR{e}{e'}{p}, if $e$ occurs in the subsequence of $h_p$ that precedes $e'$.

An \emph{execution} of a distributed system consists of
the set of all events in all process histories, together with
the \por \poR{}{}{p} over events in each $h_p$ and
the \emph{happens-before} relation \hbR{}{} over all events.
The \hbr, due to~\citet{lamport-clocks}, is an irreflexive partial order that
captures the \emph{potential causality} of events in an execution:
for any two events $e$ and $e'$,
if \hbR{e}{e'},
then $e$ may have caused $e'$,
but we can be certain that $e'$ did not cause $e$.

\begin{definition}[Happens-before (\hbR{}{})~\citep{lamport-clocks}]
    \label{def_math_happens_before}
    Given events $e$ and $e'$,
    $e$ \emph{\hb} $e'$,
    written $\hbR{e}{e'}$, iff:
    \begin{itemize}
        \item $e$ and $e'$ occur in the same process history $h_p$ and $\poR{e}{e'}{p}$; or
        \item 
            $e = \broadcastE{m}$ and
            $e' = \deliverE{p}{m}$
            for a given message $m$ and some process $p$; or
        \item $\hbR{e}{e''}$ and $\hbR{e''}{e'}$
            for some event $e''$.
    \end{itemize}
\end{definition}
\noindent
Events in the same process history are totally ordered by the \hbr
(For example, in \Cref{fig:fifo-causal-violations},
Alice's broadcast of ``I lost my wallet...'' \hb
her broadcast of ``Found it!''), and the broadcast of a given message \hb any delivery of that message.
We say that \hbR{m}{m'} iff \hbR{\broadcastE{m}}{\broadcastE{m'}}, using the notation \hbR{}{} for both relations.

To avoid executions
like those in \Cref{fig:fifo-causal-violations},
processes must deliver messages
in an order consistent with the \hbR{}{} partial order.
This property is known as \emph{\cd};
our definition is based on standard ones~\citep{raynal-causal-ordering, birman-lightweight-cbcast}:
\begin{definition}[\Cd]
    \label{def_math_cd}
    An execution $x$ observes \emph{\cd} if,
    for all processes $p$ in $x$,
    for all messages $m_1$ and $m_2$ such that
    \deliverE{p}{m_1} and
    \deliverE{p}{m_2} are in $h_p$,
    \begin{center}
        $\hbR{m_1}{m_2} \implies \poR{\deliverE{p}{m_1}}{\deliverE{p}{m_2}}{p}$.
    \end{center}
\end{definition}
\noindent
The causal delivery property says that
if message $m_1$ is sent before message $m_2$ in an execution,
then any process delivering both $m_1$ and $m_2$
should deliver $m_1$ first.
For example, in \Cref{fig:fifo-causal-violations}~(left),
the ``I lost my wallet...'' message causally precedes the ``Found it!'' message,
because Alice broadcasts both messages with ``I lost my wallet...'' first,
and so Bob and Carol would each need to deliver ``I lost my wallet...''
first for the execution to observe causal delivery.
Furthermore, under causal delivery $m_1$ and $m_2$ must be delivered
in causal order even if they were sent by different processes.
For example, in \Cref{fig:fifo-causal-violations}~(right),
Alice's ``Found it!'' message causally precedes Bob's ``Glad to hear it!'' message,
and therefore Carol, who delivers both messages,
must deliver Alice's message first for the execution to observe causal delivery.

\subsection{Background: \CBA}
\label{subsec_bg_causal_broadcast}

The \cba that we implemented and verified is due
to~\citet{birman-lightweight-cbcast}; in this section, we describe how it works at a high level before discussing our Liquid Haskell implementation in \Cref{sec_implementation}.

The protocol is based on \emph{vector clocks}, a type of logical clock well-known in the
distributed systems literature~\citep{mattern-vector-time, fidge-vector-time, schmuck-dissertation}.
Like other logical clocks, \vc{}s do not track
physical time (which would be problematic in distributed computations that lack
a global physical clock), but instead track the order of events.  Readers already familiar with vector clocks may skip ahead to \Cref{subsubsec_bg_deliverability}.

\subsubsection{\VCP}
\label{subsubsec_bg_vc}
\begin{figure}
  \includegraphics[width=0.48\columnwidth]{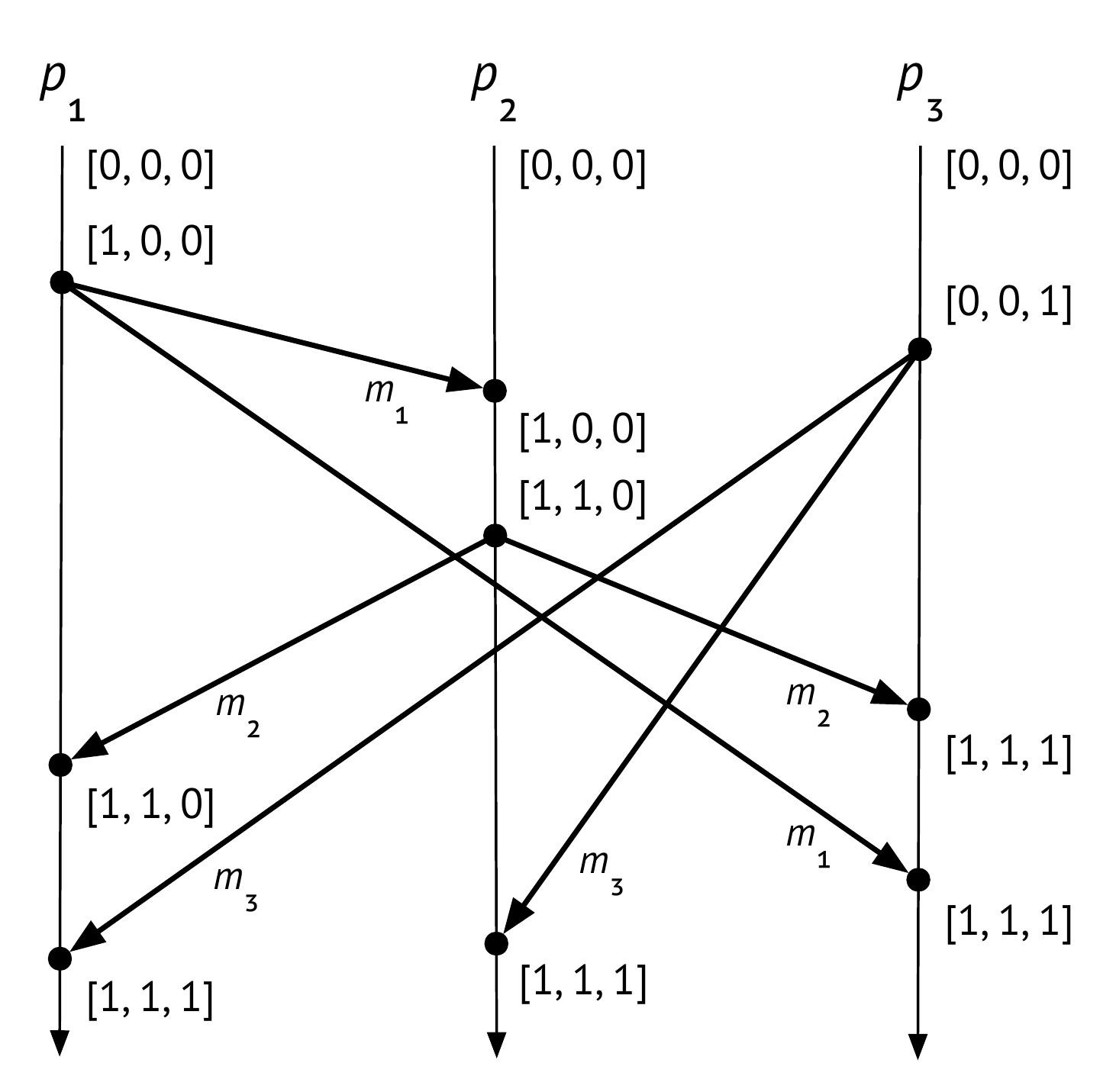}
  \caption{An example execution using the vector clock protocol.  As each
process broadcasts and delivers messages, it updates its vector clock according
to the protocol.  For example, when process $p_1$ broadcasts $m_1$, it
increments its own position in its clock just before broadcasting the
message, and $m_1$ carries the incremented clock \hs{[1,0,0]} as metadata.
  }
  \label{fig:vc-definition}
\end{figure}

A \emph{vector clock} is a sequence of length $N$ (the number of
processes in the system), which is indexed by process identifiers $i:1..N$, and
where each entry is a natural number. 
At the beginning of an execution every process $p$ initializes its own vector clock, denoted \vcF{p}, to
zeroes.
The protocol proceeds as follows:
\begin{itemize}
    \item
        When a process $p_i$ broadcasts a message $m$,
        $p_i$ increments its own position in its vector clock,
        $\vcF{p_i}[i]$, by 1.
    \item
        Each message broadcast by a process $p$
        carries as metadata the value of \vcF{p}
        that was current at the time the message was broadcast
        (just after incrementing),
        denoted \vcF{m}.
    \item
        When a process $p$ delivers a message $m$,
        $p$ updates its own vector clock \vcF{p}
        to the \emph{pointwise maximum} of \vcF{m} and \vcF{p}
        by taking the maximum of the integers at each index:
        for $k:1..N$, we update $\vcF{p}[k]$
        to $\mathrm{max}(\vcF{m}[k], \vcF{p}[k])$.
\end{itemize}
\Cref{fig:vc-definition} illustrates an example execution of three processes
running the vector clock protocol.

We can define a partial order on vector clocks of the same length as follows:
for two vector clocks $\mi{a}$ and $\mi{b}$ indexed by $i:1..N$,
\begin{itemize}
    \item $\vcleR{\mi{a}}{\mi{b}}$ if $\forall i.\, \mi{a}[i] \leq \mi{b}[i]$, and
    \item $\vcltR{\mi{a}}{\mi{b}}$ if $\vcleR{\mi{a}}{\mi{b}}$ and $\mi{a} \neq \mi{b}$. 
\end{itemize}
This ordering is not total: for example, in \Cref{fig:vc-definition},
$m_1$ carries a vector clock of @[1,0,0]@ while $m_3$ carries a vector clock
of @[0,0,1]@, and neither is less than the other.  Correspondingly, $m_1$
and $m_3$ are \emph{causally independent} (or \emph{concurrent}): neither
message has a causal dependency on the other.
On the other hand, $m_2$ causally depends on $m_1$; correspondingly,
$m_1$'s vector clock @[1,0,0]@ is less than @[1,1,0]@ carried by $m_2$.
In fact, vector clocks under this protocol \emph{precisely} characterize
the causal partial ordering~\citep{mattern-vector-time, fidge-vector-time}:
for all messages $m, m'$, it can be shown that
\begin{equation}\label{eqn:vc-hb-iff}
    \hbR{m}{m'} \iff \vcltR{\vcF{m}}{\vcF{m'}}.
\end{equation}
This powerful two-way implication lets us boil down the problem of
reasoning about causal relationships between messages in a distributed execution to
a \emph{locally checkable} property.

By itself, the \vcp does not enforce \cd of messages.
Indeed, the execution in \Cref{fig:vc-definition} violates \cd:
under \cd, process $p_3$ would not deliver $m_1$ before $m_2$.
However, the vector clock metadata attached to each message
can be used to enforce causal delivery of broadcast messages, as we will see next.

\subsubsection{Deliverability}
\label{subsubsec_bg_deliverability}

The vector clock attached to a message
can be thought of as a summary of the causal history of that message:
for example, in \Cref{fig:vc-definition},
$m_2$'s vector clock of @[1,1,0]@ expresses that one message from $p_1$
(represented by the @1@ in the first entry of the vector) causally precedes $m_2$.
Furthermore, each process's vector clock tracks how many messages
it has delivered from each process in the system.
We can exploit this property by having the recipient of each broadcast message
compare the message's attached vector clock with its own vector clock
to check for \emph{deliverability}, as follows:
\begin{definition}[Deliverability~\citep{birman-lightweight-cbcast}]
    \label{def_math_deliverable}
    A message $m$ broadcast by a process $p_i$
    is \emph{deliverable} at a process $p_j \neq p_i$
    if, for $k:1..N$,
    \begin{align*}
        \vcF{m}[k] &=    \vcF{p_j}[k] + 1 &\textrm{if $k = i$, and} \\
        \vcF{m}[k] &\leq \vcF{p_j}[k]     &\textrm{otherwise.}      
    \end{align*}
    \vspace{-0.6cm}
\end{definition}
\noindent Our notional ``mail clerk'' will use \Cref{def_math_deliverable}'s deliverability condition to decide when to deliver received messages.
How it works is a bit subtle, but worth understanding because of the key role it plays in the protocol:

\begin{itemize}
\item The first clause of \Cref{def_math_deliverable} ensures that $m$
is the recipient $p_j$'s \emph{next expected} message from the sender, $p_i$.
The number of messages from $p_i$ that $p_j$ has already delivered will appear
in \vcF{p_j} at index $i$, so \vcF{m}[i] should be \emph{exactly one
greater} than \vcF{p_j}[i].

\item The second clause ensures that $m$'s causal
history does not include any messages sent by processes \emph{other than}
$p_i$ that $p_j$ has not yet delivered.
If $m$'s vector clock is greater than
$p_j$'s vector clock in any position $k \neq i$, then it means that, before
sending $m$, process $p_i$ must have delivered some message $m'$ from $p_k$
that has not yet been delivered at $p_j$.
\end{itemize}

Combining the vector clock protocol of \Cref{subsubsec_bg_vc} with the
deliverability property of \Cref{def_math_deliverable} gives us
\citeauthor{birman-lightweight-cbcast}'s \cba.  Whenever
a process receives a message, it buffers the message until it is deliverable
according to \Cref{def_math_deliverable}.  Each process stores
messages that need to be buffered in a process-local queue, the \emph{delay
queue}.  Whenever a process delivers a
message and updates its own vector clock, it can check its delay queue for
buffered messages and deliver any messages that have become deliverable (which
may in turn make others deliverable).
\begin{figure}
  \includegraphics[width=\columnwidth]{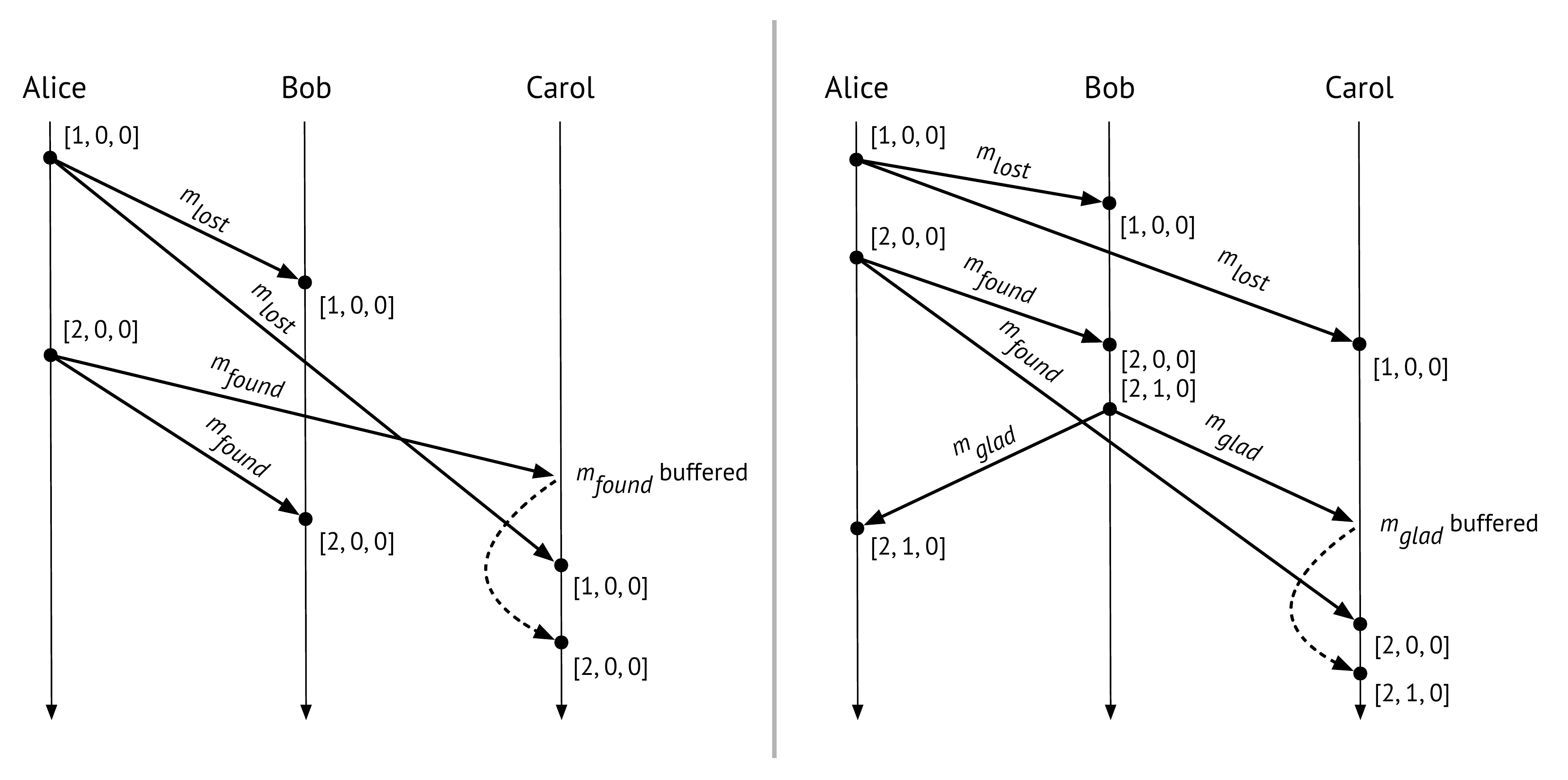}
  \caption{The executions from \Cref{fig:fifo-causal-violations},
    annotated with vector clocks used by the \cba.  On the
    left, Carol buffers $m_{\mi{found}}$ until she has delivered $m_{\mi{lost}}$.
    On the right, Carol buffers 
    $m_{\mi{glad}}$ until she has
    delivered $m_{\mi{found}}$.}
  \label{fig:vector-clocks}
\end{figure}

\subsubsection{Example Executions of the \CBA}

To illustrate how the protocol works, \Cref{fig:vector-clocks} shows the
two problematic executions we saw previously in
\Cref{fig:fifo-causal-violations}, but now with the \cba in place to prevent violations of causal delivery.
Each process keeps a vector clock with three entries corresponding to Alice,
Bob, and Carol respectively.  Suppose that $m_{\mi{lost}}$ is Alice's ``I lost
my wallet...'' message, $m_{\mi{found}}$ is Alice's ``Found it!'' message, and
$m_{\mi{glad}}$ is Bob's ``Glad to hear it!'' message.

In \Cref{fig:vector-clocks}~(left), Bob receives Alice's messages in the
order she broadcasted them, and so he can deliver them immediately.  For
example, when Bob receives $m_{\mi{lost}}$, his own vector clock is
@[0,0,0]@, and the vector clock on the message is @[1,0,0]@.  The
message is deliverable at Bob's process because it is one greater than Bob's
own vector clock in the sender's (Alice's) position, and less than or equal to
Bob's vector clock in the other positions, so Bob delivers it immediately
after receiving it.  Carol, on the other hand, receives $m_{\mi{found}}$
first.  This message has a vector clock of @[2,0,0]@, so it is not
immediately deliverable at Carol's process because Carol's vector clock is
@[0,0,0]@, and so the entry of @2@ at the sender's index is too large,
indicating that the message is ``from the future'' and needs to be buffered in
Carol's delay queue for later delivery, after Carol delivers $m_{\mi{lost}}$.

In \Cref{fig:vector-clocks}~(right), Bob delivers two messages from
Alice and then broadcasts $m_{\mi{glad}}$.  $m_{\mi{glad}}$ has a vector clock
of @[2,1,0]@, indicating that it has two messages sent by Alice in its
causal history.  When Carol receives $m_{\mi{glad}}$, her own vector clock is
only @[1,0,0]@, indicating that she has only delivered one of those
messages from Alice so far, so Carol must buffer $m_{\mi{glad}}$ in her delay
queue until she receives and delivers $m_{\mi{found}}$, the missing message
from Alice, increasing her own vector clock to @[2,0,0]@.  Now
$m_{\mi{glad}}$ is deliverable at Carol's process, and Carol can deliver it,
increasing her own vector clock to @[2,1,0]@.

\subsection{Verification Task}
\label{subsec_bg_verification_task}

Thanks to the relationship between the happens-before ordering and the vector
clock ordering expressed by \Cref{eqn:vc-hb-iff}, we can reduce the problem of determining that a distributed execution observes causal delivery to a condition that is \emph{locally} checkable at each process.  We call this condition \emph{local causal delivery}:

\begin{definition}[\Plcd]
\label{def_math_plcd}
A process $p$ observes \emph{\plcd} if, for all
    messages $m_1$ and $m_2$ such that
    \deliverE{p}{m_1} and
    \deliverE{p}{m_2}
    are in $h_p$,
    \begin{center}
      $\vcltR{\vcF{m_1}}{\vcF{m_2}}    \implies    \poR{\deliverE{p}{m_1}}{\deliverE{p}{m_2}}{p}$.
    \end{center}
\end{definition}
\noindent The heart of our verification task will be to prove that our implementation of the \cba of \Cref{subsec_bg_causal_broadcast} ensures that processes that run the protocol
observe \plcd.  From there, given \Cref{eqn:vc-hb-iff}, we can prove that executions produced by a distributed system of processes that run the \cba observe \emph{global} causal delivery:

\begin{theorem}[Global Correctness of \CBA]
\label{thm_cbcast_cd}
An execution in which all processes run the \cba observes \cd.
\end{theorem}

In the following sections, we show how we use Liquid Haskell to implement the \cba, to
make the statement of \Cref{thm_cbcast_cd} precise, and to prove \Cref{thm_cbcast_cd}.  After presenting the implementation in \Cref{sec_implementation}, in \Cref{sec_verification} we develop the machinery necessary to express \Cref{def_math_cd,def_math_plcd} and \Cref{thm_cbcast_cd} as refinement types.

\section{Implementation}
\label{sec_implementation}

In this section, we describe our implementation of \citeauthor{birman-lightweight-cbcast}'s \cba as a Liquid Haskell library.
\Cref{subsec_impl_system_model} describes the types we use to implement our system model and \vc operations,
and in \Cref{subsec_lh} we give a brief overview of refinement types and Liquid Haskell
before diving into our implementation of the protocol itself in
\Cref{subsec_impl_causal_broadcast}.
Finally, \Cref{subsec_impl_api_example} discusses how a user application would use our library.

\subsection{System Model and Vector Clocks}
\label{subsec_impl_system_model}

We begin by defining Haskell types to implement our system model and \vc operations.
Process identifiers are natural numbers and double as indexes into vector
clocks, which are represented by a list of natural numbers.
Messages have type @M r@, where the @r@ parameter is the application-defined type of the raw message content (e.g., a JSON-formatted string). 
\begin{minted}
type PID = Nat
type VC  = [Nat]
data M r = M { mVC::VC, mSender::PID, mRaw::r }
\end{minted}
A message has three fields: 
@mVC@ and @mSender@ are respectively the metadata 
that capture when the message was sent (as a @VC@)
and who sent it (as a @PID@), and 
@mRaw@ contains the raw message content.

An event can be either a @Broadcast@ (to the network) or a @Deliver@ (to the local
user application for processing), and a process history @H@ is a list of events.
\begin{minted}{haskell}
data Event r = Broadcast (M r) | Deliver PID (M r)
type H r = [Event r]
\end{minted}

To implement the vector clock protocol of \Cref{subsubsec_bg_vc}, we need some standard vector clock operations, with the below interface:

\begin{minted}{haskell}
vcEmpty     :: Nat -> VC
vcTick      :: VC -> PID -> VC
vcCombine   :: VC -> VC -> VC
vcLessEqual :: VC -> VC -> Bool
vcLess      :: VC -> VC -> Bool
\end{minted}
@vcEmpty@ initializes a \vc of a given size with zeroes, @vcTick@ increments a \vc at a given index, @vcCombine@ computes the pointwise maximum of two \vc{}s, and @vcLessEqual@ and @vcLess@ implement the vector clock ordering described in \Cref{subsubsec_bg_vc}.  As we will see in the following sections, our causal broadcast implementation uses @vcTick@ and @vcCombine@ when broadcasting and delivering messages, respectively.
The prose definitions of all these operations translate directly into idiomatic Haskell; for example, the implementation of @vcCombine@ is @zipWith max@.

\subsection{Brief Background: Refinement Types and Liquid Haskell}
\label{subsec_lh}

Traditionally, refinement types~\citep{rushby-predicate-subtyping, xi-array-dependent} have let programmers specify types augmented with logical predicates, called \emph{refinement predicates}, that restrict the set of values that can inhabit a type.
For example, in Liquid Haskell one could give @vcCombine@ the following signature:
\begin{minted}
vcCombine :: v:VC -> {v' :VC | len v'  == len v}
                  -> {v'':VC | len v'' == len v}
\end{minted}
\noindent The refinement on @v'@ expresses the precondition that @v@ and @v'@ will have the same length, and the return type expresses the postcondition that the returned vector clock will have the same length as the argument vector clocks.  Liquid Haskell automatically proves that such postconditions hold by generating verification conditions that are checked at compile time by the underlying SMT solver (by default, Z3~\cite{de-moura-z3}).  If the solver cannot ensure that the verification conditions are valid, typechecking fails.
In our actual implementation, additional Liquid Haskell refinements on @VC@ and @PID@ --- elided in this paper for readability ---
ensure that all functions are called with compatible \vc{}s (having the same length)
and @PID@s (natural numbers smaller than the length of a \vc).\footnote{
    Recall from \Cref{subsec_bg_system_model} that we model a distributed system as
    a finite set of $N$ processes.
    We want our implementation to be agnostic to $N$, yet we need to know what $N$
    is because it determines the length of \vc{}s (and hence what constitutes a
    valid index into a \vc{}).
    We accomplish this in Liquid Haskell by parameterizing types 
    with an $N$ expression value which will be provided at initialization by
    application code.
    For readability, we elide these length-indexing parameters from
    types in this paper, although they are ubiquitous in our implementation.
}

Aside from preconditions and postconditions of individual functions, though, Liquid Haskell makes it possible to verify \emph{extrinsic properties} that relate two functions, or calls to the same function applied to different inputs.  As an example, here is a Liquid Haskell proof that \hs{vcCombine} is commutative:
\begin{minted}
type Comm a A
    = x:a -> y:a -> {_:Proof | A x y == A y x}
vcCombineComm :: n:Nat -> Comm n {vcCombine}
vcCombineComm _n []      []      = ()
vcCombineComm n  (_x:xs) (_y:ys) =
  vcCombineComm (n - 1) xs ys
\end{minted}
\noindent Here, \hs{vcCombineComm} is a Haskell function that returns a value of \hs{Proof} type (a type alias for \hs{()}, Haskell's unit type), refined by the predicate \hs{vcCombine x y == vcCombine y x}.  The proof is by induction on the structure of vector clocks.  The base case, in which both \hs{x} and \hs{y} are empty lists, is automatic for the SMT solver, so the body of the base case need not say anything but \hs{()}.  The inductive case has a recursive call to \hs{vcCombineComm}.  We use a similar approach to prove that @vcCombine@ is associative, idempotent, and
inflationary, and that @vcLess@ is a strict partial order.  In general, programmers can specify arbitrary extrinsic properties in refinement types, including properties that refer to arbitrary Haskell functions via the notion of \emph{reflection}~\cite{vazou-refinement-reflection}.
The programmer can then prove those extrinsic properties by writing Haskell programs that inhabit those refinement types, using Liquid Haskell's provided \emph{proof combinators} --- with the help of the underlying SMT solver to simplify the construction of these proofs-as-programs~\cite{vazou-theorem-proving-for-all, vazou-refinement-reflection}.

Liquid Haskell thus occupies a position at the intersection of SMT-based program verifiers such as Dafny~\cite{leino-dafny}, and theorem provers that leverage the Curry-Howard correspondence such as Coq~\cite{bertot-coq} and Agda~\cite{norell-agda}.  A Liquid Haskell program can consist of both application code like \hs{vcCombine} (which runs at execution time, as usual) and verification code like \hs{vcCombineComm} (which is never run, but merely typechecked), but, pleasantly, both are just Haskell programs, albeit annotated with refinement types.  Since Liquid Haskell is based on Haskell, programmers can gradually port Haskell programs to Liquid Haskell, adding richer specifications to code as they go.
For instance, a programmer might begin with an implementation of \hs{vcCombine} with the type \hs{VC -> VC -> VC}, later refine it to the more specific refinement type above, even later prove \hs{vcCombineComm}, and still later use the proof returned by \hs{vcCombineComm} as a premise to prove another, more interesting property.

\subsection{\CBA Implementation}
\label{subsec_impl_causal_broadcast}

We express the \cba of \Cref{subsec_bg_causal_broadcast} as a state transition system.

\subsubsection{Process Type}
\label{subsubsec_process_type}

The state data structure @P r@ represents a process and is
parameterized by the type of raw content, @r@:
\begin{minted}{haskell}
data P r = P { pVC::VC, pID::PID, pDQ::[M r]
             , pHist::{ h:H r | histVC h == pVC }}
\end{minted}
The fields of @P@ include
the local vector clock @pVC@,
the local process identifier @pID@,
a delay queue of received but not-yet-delivered messages @pDQ@,
and (importantly for our verification task) the process history @pHist@.
We provide a @pEmpty :: Nat -> PID -> P r@ function that initializes a process with a \vc of the given length containing zeroes, the given process
identifier, and an empty delay queue and empty process history.

The type of the process history @pHist@ deserves further discussion, as it is our first use of a Liquid Haskell feature called \emph{datatype refinements}.  The datatype refinement on the @pHist@ field says that it contains a history @h@ of the type @H r@ defined in the previous section, but with an additional constraint 
@histVC h == pVC@.  This constraint expresses the intuition that
the vector clock @pVC@ and the history @h@ ``agree'' with each other:
for any process @p@ starting with a @pVC@ containing all zeros and an empty @pHist@,
each addition of a @Deliver (pID p) m@ event to the history for some message @m@
must coincide with an update to @pVC p@ of the form @vcCombine (mVC m) (pVC p)@.
Accordingly, @histVC h@ is defined as the supremum of \vc{}s on @Deliver@ events in @h@.
We extrinsically prove in Liquid Haskell that this @pVC@-@pHist@ agreement
property is true for the empty process and preserved by each transition in our
state transition system.
We next describe these transition functions.

\subsubsection{State Transitions}

The transition functions are @receive@, @deliver@, and @broadcast@, with the following interface:

\begin{minted}{haskell}
receive   :: M r -> P r -> P r
deliver   :: P r -> Maybe (M r, P r)
broadcast :: r -> P r -> (M r, P r)
\end{minted}
The @receive@ function adds a message from the network to the delay queue,
the @deliver@ function pops a deliverable message (if any) from the delay queue,
and the @broadcast@ function prepares raw content of type @r@ for network transport by wrapping it in a
message.
Of these transition functions, only @deliver@ and @broadcast@ are particularly interesting from
the perspective of our verification effort, 
since @receive@ only adds messages to the delay queue and cannot affect whether \cd is
violated.  We next discuss the implementation of @deliver@ and @broadcast@, respectively.

\begin{figure}
\begin{minted}{haskell}
deliver :: P r -> Maybe (M r, P r)
deliver p =
  case dequeue (pVC p) (pDQ p) of
    Nothing -> Nothing
    Just (m, pDQ') ->
    Just (m, p{ pVC = vcCombine (mVC m) (pVC p)
              , pDQ = pDQ'
              , pHist =
                  Deliver (pID p) m : pHist p })

dequeue :: VC -> DQ r -> Maybe (M r, DQ r)
dequeue _now [] = Nothing
dequeue now (x:xs)
  | deliverable x now = Just (x, xs)
  | otherwise = case dequeue now xs of
                  -- Skip past x.
                  Nothing -> Nothing
                  Just (m, xs') -> Just (m, x:xs')

deliverable :: M r -> VC -> Bool
deliverable m p_vc = let n = length p_vc in
  and (zipWith3 (deliverableHelper (mSender m))
        (finAsc n) (mVC m) p_vc)

deliverableHelper
  :: PID -> PID -> Clock -> Clock -> Bool
deliverableHelper m_id k m_vc_k p_vc_k
  | k == m_id = m_vc_k == p_vc_k + 1
  | otherwise = m_vc_k <= p_vc_k

finAsc :: n:Nat ->
  { xs:[{x:Nat | x < n}]<{\a b -> a < b}>
      | len xs == n }
\end{minted}
\caption{Implementation of \hs{deliver} and its helpers.}
\label{fig_impl_deliver}
\end{figure}

\subsubsection{Deliver}
\label{subsub_deliver}

\Cref{fig_impl_deliver} shows the implementation of @deliver@, as well as its
constituents @dequeue@, @deliverable@, and @deliverableHelper@.
At a high level, @deliver@ calls @dequeue@ on a process's delay queue and then performs bookkeeping:
If @dequeue@ popped a deliverable message,
then @deliver@ returns that message and updates the process with
a new vector clock according to the \vcp,
the new delay queue returned by @dequeue@,
and a new process history which records the delivery of the message.
The @dequeue@ function plays its part by removing and returning the first
deliverable message found in the delay queue.

The @deliverable@ predicate implements the deliverability condition of \Cref{def_math_deliverable}
to check whether a message @m@ is deliverable at time @p_vc@.
It works by calling @deliverableHelper (mSender m)@
on each offset in the message \vc @mVC m@ and process \vc @p_vc@,
and returning the conjunction of those results.
The function @finAsc n@ provides those offsets in ascending order,
and, combined with @zipWith@,
lets us implement the subtle deliverability condition of \Cref{def_math_deliverable} in @deliverableHelper@, almost exactly as
\Cref{def_math_deliverable} is written (except that our vector clocks are zero-indexed).
We omit the implementation of @finAsc@ from \Cref{fig_impl_deliver} for brevity, but its refinement type
guarantees that it returns an ascending list of length @n@ containing @Nat@s less than @n@, using Liquid Haskell's \emph{abstract refinements} feature~\citep{vazou-abstract-refinements}.

\subsubsection{Broadcast}

\begin{figure}
\begin{minted}{haskell}
broadcast :: r -> P r -> (M r, P r)
broadcast raw p =
  let m = M { mVC = vcTick (pVC p) (pID p)
            , mSender = pID p
            , mRaw = raw }
      p' = p { pDQ = m : pDQ p
             , pHist = Broadcast m : pHist p }
      Just tup = deliver p'
  in tup
\end{minted}
    \caption{Implementation of \hs{broadcast}. We prove that \hs{deliver p'} is a
    \hs{Just} value using an extrinsic proof.}
\label{fig_impl_broadcast}
\end{figure}

\Cref{fig_impl_broadcast} shows the implementation of the @broadcast@ function.
First, @broadcast@ constructs a message @m@ for the value @raw@ by incrementing
the @pID p@ index of its own \vc @pVC p@, and attaching that @pID p@ to @m@ as @mSender@.
Next, @broadcast@ constructs an intermediate process value @p'@ containing @m@ at
the head of the delay queue and a new process history recording the broadcast
event for this message.
Last, @broadcast@ delegates to @deliver@ to deliver @m@ at its own sender, @p'@.
As we will see in \Cref{sec_verification}, implementing @broadcast@ in terms of @deliver@ simplifies proving properties about our implementation, because proofs about @broadcast@ can often delegate to existing proofs about @deliver@.

Although @deliver@'s return type is @Maybe (M r, P r)@, 
the @deliver p'@ call in @broadcast@ is \emph{guaranteed} by Liquid Haskell to evaluate to
a @Just@ value containing the next process and the message to be
broadcast.
We prove this property using an extrinsic proof, not shown here.
The intuition is that messages a process sends to itself are always immediately deliverable,
because when a process increments its own index in the \vc that it places in a message, the message immediately becomes deliverable at that process.

\subsection{Example Application Architecture}
\label{subsec_impl_api_example}

The @receive@, @deliver@, and @broadcast@ functions are the interface made available to user applications of our causal broadcast library.  When @deliver@
returns a message, the user application must process it immediately.
The user application must also immediately put the message returned by
@broadcast@ on the network and also process the message locally.%
\footnote{In practical applications, it may be advantageous to separate these concerns about handling
return values into an additional message-handling layer, but that is beyond our scope.}
This design implies that user applications should not update their
own state directly when communication is in order, but rather, generate a
message and then update their state in response to its delivery.

\begin{figure}
  \includegraphics[width=\columnwidth]{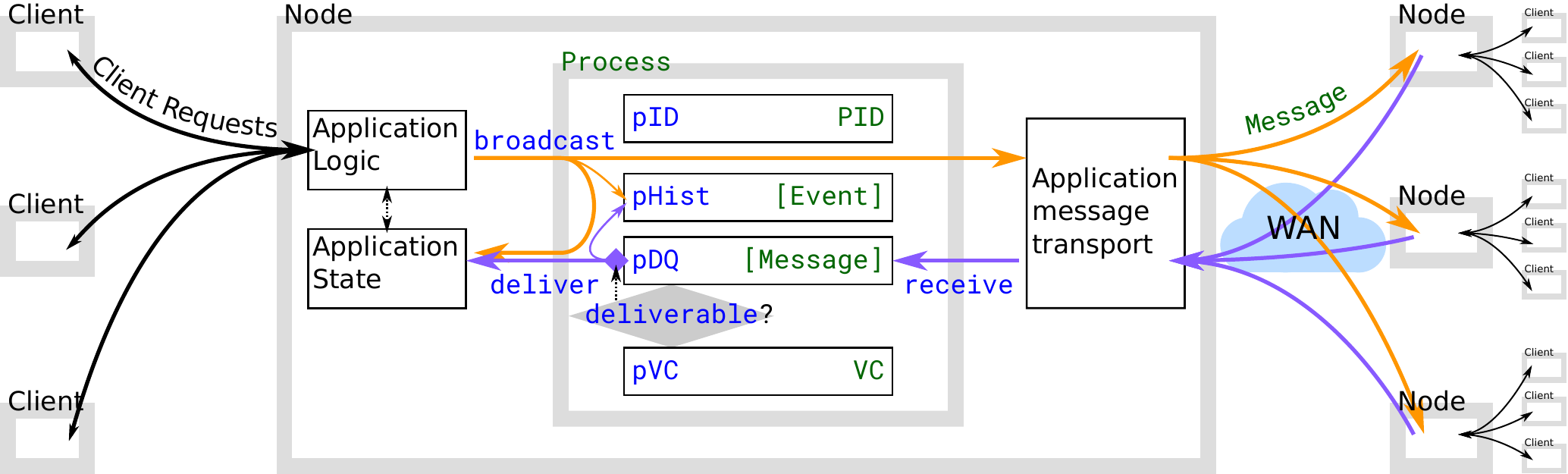}
  \caption{
      Example architecture for a distributed application using our causal
      broadcast library.
      The mnemonic standins \hs{Process}, \hs{Event}, and \hs{Message}
      refer to the types \hs{P r}, \hs{Event r}, and \hs{M r} defined in our implementation.
      An application \emph{node} using this architecture
      participates in the \cba using a single process data structure
      and the functions \hs{receive}, \hs{broadcast}, and \hs{deliver}
      to safely manage message-passing state.
      Clients make requests to a node, possibly updating application state, and
      the node may generate messages to replicate updates or perform other tasks.
    }
  \label{fig:architecture}
\end{figure}

Figure~\ref{fig:architecture} shows an example architecture of an application
using our causal broadcast library.  A collection of (potentially
geo-distributed) peer nodes, which we call the \emph{causal broadcast cluster},
each run the causal broadcast protocol along with their user application code (for
instance, a key-value store or a group chat application).
Clients of the
application communicate their requests to the nodes; one or more clients may
communicate with each node.  The application instance on a node generates
messages, broadcasts them to other nodes, and delivers messages received
from other nodes.  Later on, in \Cref{sec_demo}, we will see a case study of an application with this architecture.

\definecolor{outbound}{RGB}{255,150,  0}
\definecolor{inbound} {RGB}{136, 90,255}

\section{Verification}
\label{sec_verification}

In this section we mechanize \cd and \plcd (\Cref{def_math_plcd,def_math_cd})
for our implementation of the \cba,
and
we describe the highlights of our Liquid Haskell proof development,
culminating in a mechanized proof of \Cref{thm_cbcast_cd}.
In \Cref{subsec_verif_def_plcd} we show how we express \plcd (abbreviated ``LCD'' henceforth) as a refinement type in Liquid Haskell,
and in \Cref{subsec_verif_pres_plcd} we show that each of the
@receive@, @deliver@, and @broadcast@ transitions of \Cref{subsec_impl_causal_broadcast}
results in a process that observes LCD.
We then leverage this fact to prove \Cref{thm_cbcast_cd} in \Cref{subsec_verif_cd}.
Finally, in \Cref{subsec_verif_liveness} we briefly discuss the liveness of our implementation.

\subsection{\PLCD as a Refinement Type}
\label{subsec_verif_def_plcd}

As we saw in \Cref{subsubsec_process_type}, a process tracks the history of events that have occurred on it so far, including message broadcasts and deliveries.  We can examine a process's history and see whether the process has been delivering messages in an order consistent with the messages' vector clock ordering.  Therefore, we can express \plcd (\Cref{def_math_plcd}) as a refinement type as follows:

\begin{minted}{haskell}
type LocalCausalDelivery r ID HIST
    =  {m1 : M r | elem (Deliver ID m1) HIST }
    -> {m2 : M r | elem (Deliver ID m2) HIST
                && vcLess (mVC m1) (mVC m2) }
    -> { _:Proof | processOrder HIST
                (Deliver ID m1) (Deliver ID m2) }
\end{minted}

The type alias @LocalCausalDelivery r ID HIST@ fixes
a process identifier @ID@ and
a process history @HIST@.\footnote{
    In Liquid Haskell, type aliases can be parameterized either with ordinary
    Haskell type variables or with Liquid Haskell expression variables. In the
    latter case, the parameter is written in ALL CAPS.
}  It is the type of a function that
given messages @m1@ and @m2@,
both of which have already been delivered in the specified process history
and for which the vector clock of @m1@ is less than that of @m2@,
produces a proof that the delivery event of @m1@ precedes the delivery event of @m2@ in the process history.
The @vcLess@ function is part of the vector clock interface described in
\Cref{subsec_impl_system_model}, and the predicate @processOrder h e e'@ returns @True@ if event @e@ is present in the list of events that precede event @e'@ in a process history @h@.
  
The @LocalCausalDelivery@ type captures what it means for a given process to observe LCD: it says that if we consider any two messages that are in the process's history, and those messages' vector clocks have an order, then there is evidence -- in this case, in the form of an affirmative answer from an SMT solver -- that those messages appear in the process history in their vector clock order, rather than the other way around.  Our next step will be to show that this LCD property actually holds for processes running our implementation of the \cba.

\subsection{Local Causal Delivery Preservation}
\label{subsec_verif_pres_plcd}

Recall the state transition system consisting of the process type @P r@ and the functions @receive@, @deliver@, and @broadcast@ discussed in \Cref{subsec_impl_causal_broadcast}.
We need to prove (1) that a process observes LCD in its initial, empty state returned by @pEmpty@, and (2) that whenever a process satisfying LCD transitions to a new state via any sequence of steps of the @receive@, @deliver@, or @broadcast@ transition functions, the resulting process state still observes LCD.
A proof that the empty process observes LCD as defined in \Cref{subsec_verif_def_plcd} is trivially discharged by Liquid Haskell,
so we turn our attention to proving that each of the state transitions preserves LCD. 
Most of the action of our proof development happens in handling @deliver@ steps, as we will see below in \Cref{subsec_verif_deliver_plcd}.

To use the @LocalCausalDelivery@ type alias with the process type,
@P r@, we need a small adapter to extract the @pID@ and @pHist@ fields.\footnote{
    When instantiating a Liquid Haskell type alias parameterized by expression
    variables, the expressions are wrapped with braces to distinguish them from
    type parameters.
}
\begin{minted}{haskell}
type LCD r PROC =
    LocalCausalDelivery r {pID PROC} {pHist PROC}
\end{minted}

To encode the inputs to each of the \cba transition functions, we define a sum
type over the arguments, @Op r@.
Each function takes a @P r@ input and additional arguments corresponding to one
of the @Op r@ constructors.
\begin{minted}{haskell}
data Op r = OpReceive (M r)
          | OpDeliver
          | OpBroadcast r
\end{minted}
To apply those transition functions to a process value, we define @step@.
It branches on the constructor of @Op r@,
calls a transition function discussed in \Cref{subsec_impl_causal_broadcast},
extracts the next process value,
and throws away information unneeded for the proof.
\begin{minted}{haskell}
step :: Op r -> P r -> P r
step (OpReceive   m) p = receive m p
step (OpBroadcast r) p = case broadcast r p of
                           (_, p') -> p'
step (OpDeliver    ) p = case deliver p of
                           Just (_, p') -> p'
                           Nothing      -> p
\end{minted}
Next, we prove a @stepLCD@ lemma, which states that for a given
operation @op@ and process @p@, if LCD holds for @p@, then it still holds
after applying @op@ to @p@ using @step@:

\begin{minted}{haskell}
stepLCD :: op : Op r
        ->  p : P r
        -> LCD r {p}
        -> LCD r {step op p}
\end{minted}
The proof of @stepLCD@ branches on the constructors for @op@, followed by
delegation to lemmas about each of the transition functions.
\begin{minted}{haskell}
stepLCD op p pLCD =
  case op ? step op p of
    OpBroadcast r -> broadcastLCDpres r p pLCD
    OpReceive   m -> receiveLCDpres   m p pLCD
    OpDeliver     -> deliverLCDpres     p pLCD
\end{minted}
By far the most involved of these three lemmas is @deliverLCDpres@, the one that deals with @deliver@ steps.
Proving @broadcastLCDpres@ is straightforward because calling @broadcast@ only adds a @Broadcast@ event to the process history (and
then calls @deliver@), and so if a process observes LCD before calling
@broadcast@, then it is easy to show that it still does after adding the event (and for calling
@deliver@ to deliver the message locally, we can delegate to @deliverLCDpres@).
Proving @receiveLCDpres@ is even more straightforward because calling @receive@ does not modify the process history, and so if a process observes LCD before calling @receive@, it is easy to show that it still does afterward.
We therefore omit discussion of @receiveLCDpres@ and @broadcastLCDpres@ and focus on the proof of @deliverLCDpres@ in the next section.

\subsubsection{Deliver Transition Preservation Lemma}
\label{subsec_verif_deliver_plcd}

The @deliverLCDpres@ lemma states that a process's observation of LCD is preserved through calls to the @deliver@ function.
The proof begins by deconstructing the two cases of @dequeue@, echoing the definition of @deliver@ (\Cref{fig_impl_deliver}).
In the case that @dequeue@ returns @Nothing@,
so does its caller @deliver@,
and the process state is unchanged.
This line of reasoning is automatically carried out by Liquid Haskell
without needing to be explicitly written in the proof.
As a result, we can use the input evidence that the original process observes LCD
to complete the case.

More interesting is the case in which @dequeue@ returns a message @m@ that has been deemed deliverable.
We need to show that in an updated process state @p'@ in which @m@ has been delivered, the process still observes LCD.
Recalling the definition of @LocalCausalDelivery@ from \Cref{subsec_verif_def_plcd}, we need to show that for all messages @m1@ and @m2@ where the vector clock of @m1@ is less than that of @m2@,
@m1@'s delivery event occurs before @m2@'s delivery event in @p'@'s process history.
There are three cases to consider:

\begin{itemize}
    \item \emph{Case \hs{m == m1}}.
        When @m@ is equal to @m1@, it is the most recently delivered message on @p'@,
        but since @vcLess (mVC m1) (mVC m2)@, this would be a causal violation,
        and so we show this case is impossible.
        Recall that since @m@ was deliverable on the original process @p@, @deliverable m (pVC p)@ is @True@,
        which implies a relationship between @mVC m@ and @pVC p@:
        the @mSender m@ offset in @mVC m@ is exactly one greater than that of @pVC p@,
        and all other offsets of @mVC m@ are less than or equal to that of @pVC p@.
        Additionally,
        @vcLessEqual (mVC m1) (mVC m2)@ by @vcLess@,
        and 
        @vcLessEqual (mVC m2) (histVC p)@
            because the delivery of @m2@ is in @pHist p@ and
            because @vcCombine@ is inflationary,
        and 
        @histVC p == pVC p@ by the data refinement on processes.
        Finally, since @vcLessEqual@ is transitive, we can combine these facts to
        conclude that @vcLessEqual (mVC m1) (pVC p)@, which contradicts
        the relationship implied by @deliverable m (pVC p)@.
    \item \emph{Case \hs{m == m2}}.
        When @m@ is equal to @m2@, it is the most recently delivered message on @p'@.
        Let @e1@ be the delivery event for @m1@ with the definition @Deliver (pID p') m1@
        and similarly let @e2@ be the delivery event for the equivalent messages @m2@ and @m@.
        Since @pHist p'@ is @e2:pHist p@,
        and @e1@ is known to already be in @pHist p@,
        we can conclude that @e1@ precedes @e2@ in @p'@'s history,
        and so @processOrder (pHist p') e1 e2@, as required by LCD.
    \item \emph{Case \hs{m /= m1 && m /= m2}.}
        Finally, when @m@ is a new message distinct from both @m1@ and @m2@,
        we show that the addition of a deliver event for @m@ to @pHist p@ does not change the delivery ordering of @m1@ and @m2@.
        That is, with event @e1@ for delivery of @m1@, @e2@ for @m2@, and @e3@ for @m@,
        since @pHist p'@ is @e3:pHist p@,
        and since @e1@ and @e2@ were in @pHist p@ (and both are still in @pHist p'@),
        we can conclude that orderings about elements in @pHist p@ are unchanged in @pHist p'@.
\end{itemize}
With these pieces in place, we can conclude that a LCD-observing process continues to observe LCD after any call to @deliver@.

\begin{table}[t!]
    \centering
    \begin{tabular}{p{0.8\columnwidth} r}
      \toprule
      \textbf{Description} & \textbf{LOC}  \\
      \midrule
      Implementation without refinements                                          & 236   \\
      Implementation-supporting proofs and refinements                            & 448   \\
      \midrule
      List lemmas, extra proof combinators, shims                                 & 161   \\
      Proofs about relations           (\Cref{subsec_impl_system_model})          & 217   \\
      Model for preservation of LCD    (\Cref{subsec_verif_def_plcd})             & 27    \\
      LCD preservation                 (\Cref{{subsec_verif_pres_plcd}})          & 51    \\
      LCD preservation, @broadcast@ case                                          & 64    \\
      LCD preservation, @receive@ case                                             & 44    \\
      LCD preservation, @deliver@ case (\Cref{subsec_verif_deliver_plcd})         & 273   \\
      Model for preservation of CD     (\Cref{subsec_verif_cd})                   & 130   \\
      CD preservation                                                             & 138   \\
      CD preservation via LCD                                                     & 139   \\

      \bottomrule
    \end{tabular}
    \caption{Lines of code used in our implementation and proof development.  The LOC count includes Liquid Haskell definitions, theorems, proofs, and other annotations.}
    \label{table_proofs}
\end{table}

\subsection{Global Causal Delivery Preservation}
\label{subsec_verif_cd}

The @lcdStep@ property we proved in the previous section says that running the \cba for one step on a given process preserves \plcd for that process.  However, \Cref{thm_cbcast_cd} pertains to entire executions as opposed to individual processes.  To complete the proof, then, we must define an additional \emph{global} state transition system, where states represent executions, and a step nondeterministically picks any process in an execution and runs the \cba for one (local) step on that process.  Unlike the local state transition system, which is actually what is used at run time to execute the \cba, our global states and global steps are for verification purposes only.

We define a global execution state as a mapping from @PID@s to @P r@ process states.  We can then express (global) causal delivery (\Cref{def_math_cd}) as a refinement type, as follows:

\pagebreak
\begin{minted}{haskell}
type CausalDelivery r X
  =  pid : PID
  -> {m1 : M r | elem (Deliver pid m1)
                      (pHist (X pid)) }
  -> {m2 : M r | elem (Deliver pid m2)
                      (pHist (X pid))
                   && happensBefore X
                        (Broadcast m1)
                        (Broadcast m2) }
  -> {_: Proof | processOrder (pHist (X pid))
                   (Deliver pid m1)
                   (Deliver pid m2) }
\end{minted}
The @CausalDelivery@ type is reminiscent of the @LocalCausalDelivery@ type that we saw in \Cref{subsec_verif_def_plcd}, but instead of referring to one particular process, it refers to an entire execution, @X@.  @CausalDelivery r X@ says that for any process in @X@, messages are delivered in causal order on that process.  Another key difference is that instead of using @vcLess@, @CausalDelivery@ uses a @happensBefore@ predicate, which takes an execution argument and two events.  This is as it should be; the \emph{definition} of causal delivery should be agnostic to the \emph{mechanism} used by our particular protocol.  However, our @lcdStep@ lemma only establishes that messages on a process are delivered in an order consistent with their vector-clock ordering, not the happens-before ordering.  To bridge this gap and get from \plcd to global causal delivery, we must leverage \Cref{eqn:vc-hb-iff}'s correspondence between vector clocks and happens-before, which we express as a pair of \emph{axioms} in Liquid Haskell, one for each direction of the correspondence.

We can now prove that a single \emph{global} execution step preserves causal delivery.  The @xStepCD@ lemma states that if we have a causal-delivery-observing execution @x@, if we pick out any given process (identified by @pid@) from that execution and run any given operation @op@ on that process, then the resulting execution will also observe causal delivery.

\begin{minted}{haskell}
xStepCD :: op: Op r
        -> x: Execution
        -> pid: PID
        -> CausalDelivery r x
        -> CausalDelivery r {xStep op pid x}
\end{minted}
The proof of @xStepCD@ proceeds in three stages:

\begin{enumerate}
  \item \emph{Global to local.} First, we show that if the original execution observes causal delivery, then every process in it observes local causal delivery.  For this, we use the \emph{reflection} direction of the vector-clock/happens-before correspondence, which says that messages with a given vector clock ordering were broadcast in the corresponding happens-before order.
  \item \emph{Local step.} Next, we show that if any process in an execution takes a local step, then every process in the execution will still observe local causal delivery.  This is easy to show using our @lcdStep@ lemma.
  \item \emph{Local to global.} Finally, we show that if every process in an execution observes local causal delivery, then the entire execution observes causal delivery.  For this, we use the \emph{preservation} direction of the vector-clock/happens-before correspondence, which says that messages broadcast in a given happens-before order have the corresponding vector clock order.
\end{enumerate}
Since the vector-clock/happens-before correspondence lets us reason in a \emph{process-local} fashion, instead of having to reason about events spread across a global execution using happens-before, we enjoy a sort of ``local reasoning for free'' without the need for a more heavyweight proof technique such as separation logic.  With the proof of @xStepCD@ complete, all that remains to prove \Cref{thm_cbcast_cd} is to extrapolate from global executions that take one step to those that take any number of steps, which is straightforward to do in Liquid Haskell by induction on the number of steps.  Since an empty global execution observes causal delivery, we can conclude that any global execution where all processes are running our protocol observes causal delivery, completing the proof of \Cref{thm_cbcast_cd}.

\begin{figure}
  \includegraphics[width=\columnwidth]{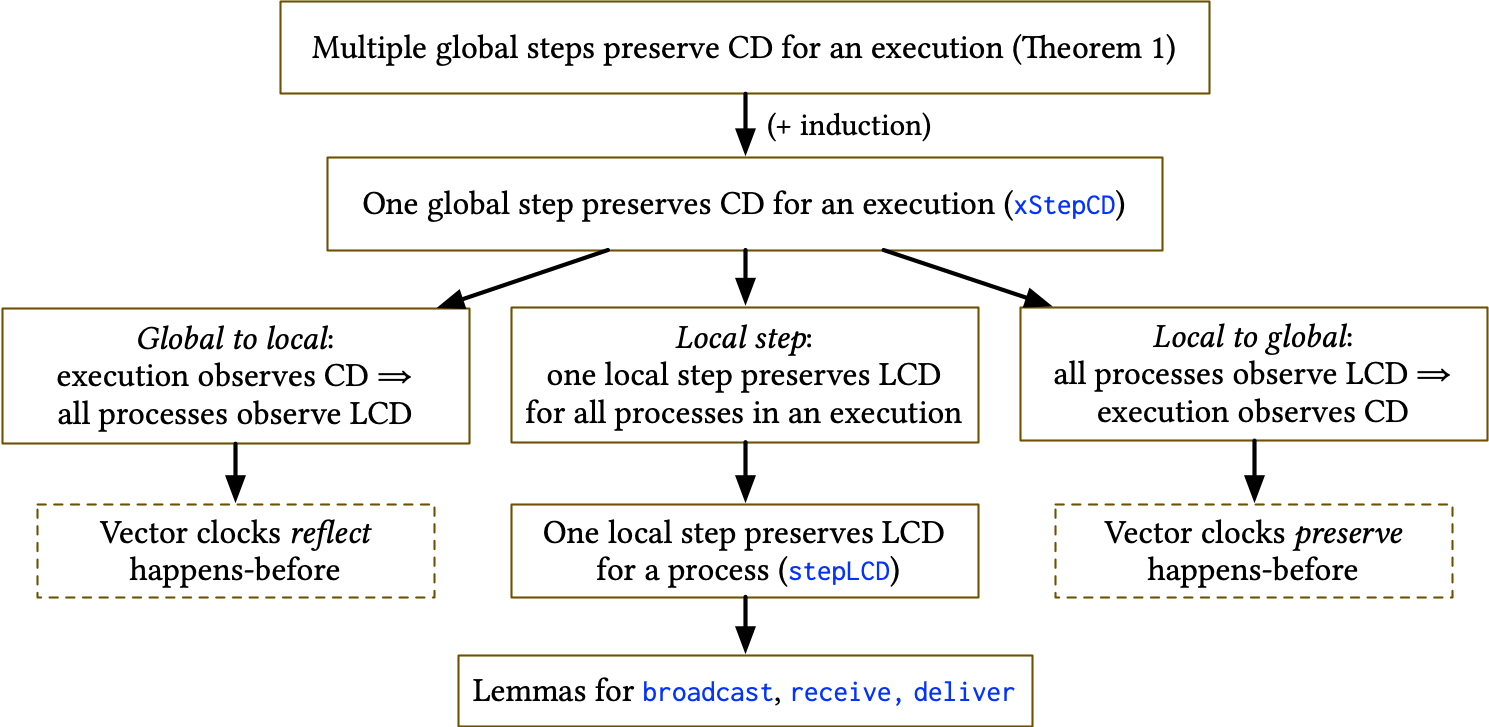}
  \caption{A high-level overview of the key components of our proof development.  Arrows indicate dependencies, solid boxes indicate theorems and lemmas, and dashed boxes indicate axioms.}
  \label{fig:proof-overview}
\end{figure}

\Cref{table_proofs} summarizes the size of each component of our proof development in terms of lines of Liquid Haskell code, and \Cref{fig:proof-overview} gives a visual overview of the important components of the proof: the @xStepCD@ property and its proof in three stages outlined above; the @stepLCD@ property and its reliance on lemmas for @broadcast@, @receive@, and @deliver@, and our use of the two directions of the vector-clock/happens-before correspondence.  In all, our proof development weighs in at 1692 lines of code for 236 lines of implementation code.

\subsection{Discussion: Liveness}
\label{subsec_verif_liveness}

A useful implementation is not only safe, but \emph{live}, which in our case would mean that messages
will not languish forever in the delay queue.
As mentioned in \Cref{subsec_bg_system_model},
for our safety result we need not make any assumption of reliable message receipt,
since we do not have to worry about the delivery order of messages that are never received.
A proof of liveness, though, would need to rest on the assumption of a reliable message transport layer,
that is, one in which sent messages are eventually received ---
albeit in arbitrary order and with arbitrarily long latency.
Otherwise, a message could be stuck forever in the delay queue if a message that causally precedes it is lost,
because it would never become deliverable.
Proofs of liveness properties are considered ``much harder''~\citep{hawblitzel-ironfleet} than proofs of safety properties.  While we do not offer any mechanized liveness proof, in the following section we argue informally for the liveness of our implementation under the reliable message reception assumption.

\section{Case study}
\label{sec_demo}

In this section we describe a key-value store (KVS) application
implemented in the architectural pattern depicted by \Cref{fig:architecture}
and using our causal broadcast library from \Cref{sec_implementation}.
The KVS is an in-memory replicated data store consisting of
message-passing nodes, each of which simultaneously serves client requests via
HTTP.
\Cref{subsec_demo_impl} covers the implementation of the KVS and demonstrates
that it is not difficult to integrate our \cba with an application to obtain
the benefits of causal broadcast.  In particular, causal broadcast can be used to ensure \emph{causal consistency} of replicated data~\citep{ahamad-causal-memory, lloyd-cops}.\footnote{
    For simplicity, we adopt a ``sticky sessions''
    model, in which a given client will only ever talk to a given server.  In a
    setting where clients can communicate with more than one server, clients
    would need to participate in the propagation of causal metadata generated
    by the servers~\citep{lloyd-cops}, whereas with sticky sessions, causal metadata is only exchanged among the servers.
} 
In \Cref{subsec_demo_eval} we describe how we deployed the KVS to a cluster of
geo-distributed nodes and evaluated its performance.

\subsection{Design and Implementation}
\label{subsec_demo_impl}

We implemented the KVS using several commonly used Haskell libraries,
such as 
\texttt{servant}~\citep{mestanogullari-servant} to express HTTP endpoints concisely as types,
\texttt{stm} to express multithreaded access to state,
\texttt{ekg} to gather runtime statistics,
and \texttt{aeson} to provide JSON (de)serialization.
Clients may request to
\texttt{PUT} a value at a key,
\texttt{DELETE} a key-value pair specified by key,
or \texttt{GET} the value corresponding to a specified key.
Servers broadcast by directly \texttt{POST}ing messages to each other.
Nodes receiving \texttt{PUT} or \texttt{DELETE} requests from their local clients call @broadcast@ to prepare a message to be immediately applied locally and broadcast to other nodes.  When a node makes a \texttt{POST} request, the endpoint calls \hs{receive} to inject the message into the node's delay queue.   Changes to the delay queue wake a background thread which calls \hs{deliver}, possibly removing a message from the delay queue and applying it to the process state.
Since messages received via the \texttt{POST} endpoint are from other
nodes, \hs{deliver} will return \hs{Nothing} in cases where the causal
dependencies of the message are not satisfied.
Therefore all nodes (and hence all clients of those nodes) observe the effects
of causally-related \hs{KvCommand}s in the same (causal) order.

\subsection{Deployment and Evaluation}
\label{subsec_demo_eval}

We deployed an eight-node KVS causal broadcast cluster,
globally distributed across AWS regions
(two nodes in \emph{us-west-1} (N. California),
 one in \emph{us-west-2} (Oregon),
 two in \emph{us-east-1} (N. Virginia),
 one in \emph{ap-northeast-1} (Tokyo), and
 two in \emph{eu-central-1} (Frankfurt)),
and 24 client nodes with three clients assigned to each KVS node.
All the nodes were AWS EC2 \texttt{t3.micro} instances with 2 vCPUs at 2.5 GHz
and 1 GiB of memory.
The 50th-percentile inter-region ping latencies vary from about 20ms between us-west-1 and us-west-2 to about 225ms between ap-northeast-1 and eu-central-1.
Each of the eight nodes in the cluster ran an instance of our
KVS application compiled with GHC 8.10.7.

We conducted a simple experiment in which each of the 24 clients made 10,000 \texttt{curl} requests
at 20 requests per second
to their assigned KVS replica in the same region
(for a total of 240,000 client requests),
uniformly distributed over \texttt{GET}, \texttt{PUT}, and \texttt{DELETE} requests.
For \texttt{PUT} requests, we used randomly generated JSON data for values,
and ensured that there were key collisions,
requiring resolution by causal order,
by drawing keys from among the
lowercase ASCII characters.

Two-thirds (160,000) of the 240,000 requests generated by clients were
\texttt{PUT} and \texttt{DELETE} requests.
Each resulted in a broadcast from the client's assigned KVS replica to the seven other nodes in the cluster,
generating 160,000 $\times$ 7 = 1,120,000 unicast messages among the eight KVS nodes.
To alleviate this message amplification and maintain throughput, we sent multiple unicast
messages in each request; typically, two or three messages were sent at a time.
The KVS replicas handled all requests and delivered all messages in the time it took for clients to send
them (10 minutes) with a load average of 0.10,
indicating that the cluster was not CPU-bound
and that no messages got stuck indefinitely in delay queues.
As a static verification approach,
Liquid Haskell itself imposes no running time overhead compared to vanilla
Haskell, and no Liquid Haskell annotations were required in the KVS application
code.

We recorded the length of the delay queue after each message delivery and
maintained an average.
Over all nodes, the average length of the delay queue after a delivery came
to 7.2 delayed messages.
From prior experiments with a different mix of KVS nodes and clients, we
observe that more nodes in the causal broadcast cluster results in
increased likelihood of messages being received out of causal order,
motivating the need for causal broadcast.

\section{Related Work}
\label{sec_related_work}

\paragraph{Machine-checked correctness proofs of executable distributed protocol implementations.}

Much work on distributed systems verification has focused on specifying and verifying properties of models using tools such as TLA+~\citep{lamport-tla}, rather than of executable implementations. Here, our focus is on mechanized verification of executable distributed protocol implementations; lacking space for a comprehensive account of the literature, we mention a few highlights.

Verdi~\citep{wilcox-verdi} is a Coq framework for implementing distributed systems; verified executable OCaml implementations can be extracted from Coq.  IronFleet~\citep{hawblitzel-ironfleet} uses the Dafny verification language, which compiles both to verification conditions checked by an SMT solver and to executable code.  Both Verdi and IronFleet have been used to verify safety properties (in particular, linearizability) of distributed consensus protocol implementations (Raft and Multi-Paxos, respectively) and of strongly-consistent key-value store implementations, and IronFleet additionally considers liveness properties.  The ShadowDB project~\citep{schiper-shadowdb} uses a language called EventML
that compiles both to a logical specification and to executable code that is automatically guaranteed to satisfy the specification, and correctness properties of the logical specification can then be proved using the Nuprl proof assistant.  \citet{schiper-shadowdb} used this workflow to verify the correctness of a Paxos-based atomic broadcast protocol.  None of \citeauthor{wilcox-verdi}, \citeauthor{hawblitzel-ironfleet}, or \citeauthor{schiper-shadowdb} looked at causal broadcast or causal message ordering in particular.

\citet{lesani-chapar} present a technique and Coq-based framework for mechanically verifying the causal consistency of distributed key-value store (KVS) implementations, with executable OCaml KVSes extracted from Coq.  \citeauthor{lesani-chapar}'s verification approach effectively bakes a notion of causal message delivery into an abstract causal operational semantics that specifies how a causally consistent KVS should behave.  In more recent work, \citet{gondelman-causal-consistency-aneris} use the Coq-based Aneris distributed separation logic framework~\citep{krogh-jespersen-aneris} --- itself built on top of the Iris separation logic framework~\citep{jung-iris} --- to specify and verify the causal consistency of a distributed KVS and further verify the correctness of a session manager library implemented on top of the KVS.  These implementations are written in AnerisLang, a domain-specific language intended to be used with the Aneris framework for implementing distributed systems.
Both \citeauthor{lesani-chapar}'s and \citeauthor{gondelman-causal-consistency-aneris}'s work is specific to the KVS use case, whereas our verified causal broadcast implementation factors out causal message delivery into a separate layer, agnostic to the content of messages, that can be used as a standalone component in a variety of applications.
Moreover, Liquid Haskell's SMT automation simplifies our proof effort by comparison.  Unlike \citeauthor{lesani-chapar} and \citeauthor{gondelman-causal-consistency-aneris}, we did not attempt to verify the causal consistency of our KVS.  However, we hypothesize that building on an underlying verified causal messaging layer would simplify the KVS verification task by separating lower-level message delivery concerns from higher-level application semantics.

\paragraph{Causal broadcast for CRDT convergence.} Conflict-free replicated data types (CRDTs)~\citep{shapiro-crdts, shapiro-crdts-comprehensive} are data structures designed for replication.  Their operations must satisfy certain mathematical properties that can be leveraged to ensure \emph{strong convergence}~\citep{shapiro-crdts}, meaning that replicas are guaranteed to have equivalent state if they have received and applied the same unordered set of updates.  While the simplest CRDTs ask little of the underlying messaging layer,
many CRDTs
implemented in the \emph{operation-based} style
rely on causal delivery to ensure that, for example, a message updating an element of a set will not be delivered before the message inserting that element.

\citet{gomes-verifying-sec} use the Isabelle/HOL proof assistant~\cite{wenzel-isabelle} to implement and verify the strong convergence of operation-based CRDTs under an assumption of causal delivery, modeled by the network axioms in their proof development.
Our work is complementary to \citeauthor{gomes-verifying-sec}'s: one could deploy their verified-convergent CRDTs atop our verified causal broadcast protocol to get an ``end-to-end'' convergence guarantee on top of a weaker network model that offers no causal delivery guarantee.

\citet{liu-verified-rdts} use Liquid Haskell to verify the convergence of operation-based CRDT implementations. \citeauthor{liu-verified-rdts}'s CRDTs do \emph{not} assume causal delivery, which complicates their implementation (and verification).  In fact, \citeauthor{liu-verified-rdts}'s verified two-phase map implementation includes a ``pending buffer'' for updates that arrived out of order, and a collection of data-structure-specific rules to determine which updates should be buffered.  These mechanisms resemble the delay queue and the \hs{deliverable} predicate, but are specific to application-level data structures and use an ad hoc delivery policy, rather than operating at the messaging layer and using the more general principle of causal delivery.  We hypothesize that our library could lessen the need for such ad hoc mechanisms.

The most closely related work to this paper --- and the only other mechanically verified causal broadcast implementation that we are aware of --- was recently carried out by \citet{nieto-crdts-aneris} as part of a larger proof development that verifies the correctness of a variety of CRDTs using the aforementioned Aneris separation logic framework.  \citeauthor{nieto-crdts-aneris}'s proof development consists of a verified stack of components, at the base of which is a verified causal broadcast library, followed by a library of CRDT components, and finally CRDT implementations.  To verify the causal broadcast library, \citeauthor{nieto-crdts-aneris} take a similar approach to \citeauthor{gondelman-causal-consistency-aneris}'s aforementioned verified key-value store, but adapted to the more general setting of causal broadcast. Their approach thus supports our hypothesis that it is possible to simplify the verification of higher-level application properties, such as causal consistency of a key-value store or convergence of CRDTs, by decoupling them from lower-level message delivery properties, such as causal broadcast.

Compared to our work, \citeauthor{nieto-crdts-aneris}'s verification effort is more broadly scoped: most obviously, they tackle verification of \emph{clients} of causal broadcast, in addition to the causal broadcast protocol itself.  Additionally, their implementation is intended to be used on top of an unreliable transport protocol, UDP, and as such it includes mechanisms to ensure reliable message delivery (although their verification, like ours, is limited to safety properties only).\footnote{Our own protocol implementation also makes no assumptions about the reliability of the underlying transport layer, but it has no mechanisms to ensure reliable delivery itself, so users of our library who do require reliable delivery should opt for a transport protocol such as TCP that provides reliable delivery out of the box.}  We deploy and empirically evaluate the performance of our implementation, whereas \citeauthor{nieto-crdts-aneris} do not.
Finally, our approach differs from \citeauthor{nieto-crdts-aneris}'s conceptually in that we frame the problem in terms of refinement types, whereas \citeauthor{nieto-crdts-aneris} take the separation-logic approach of defining logical resources and giving specifications about how those resources are used by their implementations.  Our use of Liquid Haskell lets us take advantage of SMT automation where possible, using manual proofs only when needed.  On the other hand, \citeauthor{nieto-crdts-aneris}'s use of standard separation logic mechanisms is a boon for modularity.

\section{Conclusion}
\label{sec_conclusion}

Causal message broadcast is a widely used building block of distributed applications, motivating the need for practically usable verified implementations.  We use Liquid Haskell to give a novel encoding of causal message delivery as a refinement type.  We then verify the safety of an executable causal broadcast library implemented in Haskell using a combination of manual theorem proving and SMT automation.  Our verified-safe library can be used in real distributed systems, as we demonstrate with a case-study implementation and deployment of a distributed key-value store.

\paragraph{Acknowledgments}

This material is based upon work supported by the National Science Foundation under Grant No. CCF-2145367. Any opinions, findings, and conclusions or recommendations expressed in this material are those of the author(s) and do not necessarily reflect the views of the National Science Foundation.

\newpage

\bibliography{refs}


\begin{thebibliography}{42}


\ifx \showCODEN    \undefined \def \showCODEN     #1{\unskip}     \fi
\ifx \showDOI      \undefined \def \showDOI       #1{#1}\fi
\ifx \showISBNx    \undefined \def \showISBNx     #1{\unskip}     \fi
\ifx \showISBNxiii \undefined \def \showISBNxiii  #1{\unskip}     \fi
\ifx \showISSN     \undefined \def \showISSN      #1{\unskip}     \fi
\ifx \showLCCN     \undefined \def \showLCCN      #1{\unskip}     \fi
\ifx \shownote     \undefined \def \shownote      #1{#1}          \fi
\ifx \showarticletitle \undefined \def \showarticletitle #1{#1}   \fi
\ifx \showURL      \undefined \def \showURL       {\relax}        \fi
\providecommand\bibfield[2]{#2}
\providecommand\bibinfo[2]{#2}
\providecommand\natexlab[1]{#1}
\providecommand\showeprint[2][]{arXiv:#2}

\bibitem[Acharya and Badrinath(1992)]%
        {acharya-causal-snapshots}
\bibfield{author}{\bibinfo{person}{Arup Acharya} {and} \bibinfo{person}{B.R.
  Badrinath}.} \bibinfo{year}{1992}\natexlab{}.
\newblock \showarticletitle{Recording distributed snapshots based on causal
  order of message delivery}.
\newblock \bibinfo{journal}{\emph{Inform. Process. Lett.}}
  \bibinfo{volume}{44}, \bibinfo{number}{6} (\bibinfo{year}{1992}),
  \bibinfo{pages}{317 -- 321}.
\newblock
\showISSN{0020-0190}
\urldef\tempurl%
\url{https://doi.org/10.1016/0020-0190(92)90107-7}
\showDOI{\tempurl}


\bibitem[Ahamad et~al\mbox{.}(1995)]%
        {ahamad-causal-memory}
\bibfield{author}{\bibinfo{person}{Mustaque Ahamad}, \bibinfo{person}{Gil
  Neiger}, \bibinfo{person}{James~E. Burns}, \bibinfo{person}{Prince Kohli},
  {and} \bibinfo{person}{Phillip~W. Hutto}.} \bibinfo{year}{1995}\natexlab{}.
\newblock \showarticletitle{Causal memory: definitions, implementation, and
  programming}.
\newblock \bibinfo{journal}{\emph{Distributed Computing}} \bibinfo{volume}{9},
  \bibinfo{number}{1} (\bibinfo{year}{1995}), \bibinfo{pages}{37--49}.
\newblock
\showISBNx{1432-0452}
\urldef\tempurl%
\url{https://doi.org/10.1007/BF01784241}
\showDOI{\tempurl}


\bibitem[Alagar and Venkatesan(1994)]%
        {alagar-causal-snapshots}
\bibfield{author}{\bibinfo{person}{Sridhar Alagar} {and} \bibinfo{person}{S.
  Venkatesan}.} \bibinfo{year}{1994}\natexlab{}.
\newblock \showarticletitle{An optimal algorithm for distributed snapshots with
  causal message ordering}.
\newblock \bibinfo{journal}{\emph{Inform. Process. Lett.}}
  \bibinfo{volume}{50}, \bibinfo{number}{6} (\bibinfo{year}{1994}),
  \bibinfo{pages}{311 -- 316}.
\newblock
\showISSN{0020-0190}
\urldef\tempurl%
\url{https://doi.org/10.1016/0020-0190(94)00055-7}
\showDOI{\tempurl}


\bibitem[Bertot and Cast{\'{e}}ran(2004)]%
        {bertot-coq}
\bibfield{author}{\bibinfo{person}{Yves Bertot} {and} \bibinfo{person}{Pierre
  Cast{\'{e}}ran}.} \bibinfo{year}{2004}\natexlab{}.
\newblock \bibinfo{booktitle}{\emph{Interactive Theorem Proving and Program
  Development - Coq'Art: The Calculus of Inductive Constructions}}.
\newblock \bibinfo{publisher}{Springer}.
\newblock
\showISBNx{978-3-642-05880-6}
\urldef\tempurl%
\url{https://doi.org/10.1007/978-3-662-07964-5}
\showDOI{\tempurl}


\bibitem[Birman and Joseph(1987a)]%
        {birman-virtual-synchrony}
\bibfield{author}{\bibinfo{person}{K. Birman} {and} \bibinfo{person}{T.
  Joseph}.} \bibinfo{year}{1987}\natexlab{a}.
\newblock \showarticletitle{Exploiting Virtual Synchrony in Distributed
  Systems}.
\newblock \bibinfo{journal}{\emph{SIGOPS Oper. Syst. Rev.}}
  \bibinfo{volume}{21}, \bibinfo{number}{5} (\bibinfo{date}{Nov.}
  \bibinfo{year}{1987}), \bibinfo{pages}{123–138}.
\newblock
\showISSN{0163-5980}
\urldef\tempurl%
\url{https://doi.org/10.1145/37499.37515}
\showDOI{\tempurl}


\bibitem[Birman et~al\mbox{.}(1991)]%
        {birman-lightweight-cbcast}
\bibfield{author}{\bibinfo{person}{Kenneth Birman}, \bibinfo{person}{Andr\'{e}
  Schiper}, {and} \bibinfo{person}{Pat Stephenson}.}
  \bibinfo{year}{1991}\natexlab{}.
\newblock \showarticletitle{Lightweight Causal and Atomic Group Multicast}.
\newblock \bibinfo{journal}{\emph{ACM Trans. Comput. Syst.}}
  \bibinfo{volume}{9}, \bibinfo{number}{3} (\bibinfo{date}{Aug.}
  \bibinfo{year}{1991}), \bibinfo{pages}{272–314}.
\newblock
\showISSN{0734-2071}
\urldef\tempurl%
\url{https://doi.org/10.1145/128738.128742}
\showDOI{\tempurl}


\bibitem[Birman and Joseph(1987b)]%
        {birman-reliable}
\bibfield{author}{\bibinfo{person}{Kenneth~P. Birman} {and}
  \bibinfo{person}{Thomas~A. Joseph}.} \bibinfo{year}{1987}\natexlab{b}.
\newblock \showarticletitle{Reliable Communication in the Presence of
  Failures}.
\newblock \bibinfo{journal}{\emph{ACM Trans. Comput. Syst.}}
  \bibinfo{volume}{5}, \bibinfo{number}{1} (\bibinfo{date}{Jan.}
  \bibinfo{year}{1987}), \bibinfo{pages}{47–76}.
\newblock
\showISSN{0734-2071}
\urldef\tempurl%
\url{https://doi.org/10.1145/7351.7478}
\showDOI{\tempurl}


\bibitem[Bouajjani et~al\mbox{.}(2017)]%
        {bouajjani-verifying-cc}
\bibfield{author}{\bibinfo{person}{Ahmed Bouajjani},
  \bibinfo{person}{Constantin Enea}, \bibinfo{person}{Rachid Guerraoui}, {and}
  \bibinfo{person}{Jad Hamza}.} \bibinfo{year}{2017}\natexlab{}.
\newblock \showarticletitle{On Verifying Causal Consistency}. In
  \bibinfo{booktitle}{\emph{Proceedings of the 44th ACM SIGPLAN Symposium on
  Principles of Programming Languages}} (Paris, France)
  \emph{(\bibinfo{series}{POPL 2017})}. \bibinfo{publisher}{Association for
  Computing Machinery}, \bibinfo{address}{New York, NY, USA},
  \bibinfo{pages}{626–638}.
\newblock
\showISBNx{9781450346603}
\urldef\tempurl%
\url{https://doi.org/10.1145/3009837.3009888}
\showDOI{\tempurl}


\bibitem[de~Moura and Bj{\o}rner(2008)]%
        {de-moura-z3}
\bibfield{author}{\bibinfo{person}{Leonardo de Moura} {and}
  \bibinfo{person}{Nikolaj Bj{\o}rner}.} \bibinfo{year}{2008}\natexlab{}.
\newblock \showarticletitle{Z3: An Efficient SMT Solver}. In
  \bibinfo{booktitle}{\emph{Tools and Algorithms for the Construction and
  Analysis of Systems}}, \bibfield{editor}{\bibinfo{person}{C.~R. Ramakrishnan}
  {and} \bibinfo{person}{Jakob Rehof}} (Eds.). \bibinfo{publisher}{Springer
  Berlin Heidelberg}, \bibinfo{address}{Berlin, Heidelberg},
  \bibinfo{pages}{337--340}.
\newblock
\showISBNx{978-3-540-78800-3}


\bibitem[Fidge(1988)]%
        {fidge-vector-time}
\bibfield{author}{\bibinfo{person}{C.~J. Fidge}.}
  \bibinfo{year}{1988}\natexlab{}.
\newblock \showarticletitle{Timestamps in message-passing systems that preserve
  the partial ordering}.
\newblock \bibinfo{journal}{\emph{Proceedings of the 11th Australian Computer
  Science Conference}} \bibinfo{volume}{10}, \bibinfo{number}{1}
  (\bibinfo{year}{1988}), \bibinfo{pages}{56–66}.
\newblock


\bibitem[Gomes et~al\mbox{.}(2017)]%
        {gomes-verifying-sec}
\bibfield{author}{\bibinfo{person}{Victor B.~F. Gomes}, \bibinfo{person}{Martin
  Kleppmann}, \bibinfo{person}{Dominic~P. Mulligan}, {and}
  \bibinfo{person}{Alastair~R. Beresford}.} \bibinfo{year}{2017}\natexlab{}.
\newblock \showarticletitle{Verifying Strong Eventual Consistency in
  Distributed Systems}.
\newblock \bibinfo{journal}{\emph{Proc. ACM Program. Lang.}}
  \bibinfo{volume}{1}, \bibinfo{number}{OOPSLA}, Article
  \bibinfo{articleno}{109} (\bibinfo{date}{Oct.} \bibinfo{year}{2017}),
  \bibinfo{numpages}{28}~pages.
\newblock
\urldef\tempurl%
\url{https://doi.org/10.1145/3133933}
\showDOI{\tempurl}


\bibitem[Gondelman et~al\mbox{.}(2021)]%
        {gondelman-causal-consistency-aneris}
\bibfield{author}{\bibinfo{person}{L\'{e}on Gondelman},
  \bibinfo{person}{Simon~Oddershede Gregersen}, \bibinfo{person}{Abel Nieto},
  \bibinfo{person}{Amin Timany}, {and} \bibinfo{person}{Lars Birkedal}.}
  \bibinfo{year}{2021}\natexlab{}.
\newblock \showarticletitle{Distributed Causal Memory: Modular Specification
  and Verification in Higher-Order Distributed Separation Logic}.
\newblock \bibinfo{journal}{\emph{Proc. ACM Program. Lang.}}
  \bibinfo{volume}{5}, \bibinfo{number}{POPL}, Article \bibinfo{articleno}{42}
  (\bibinfo{date}{Jan.} \bibinfo{year}{2021}), \bibinfo{numpages}{29}~pages.
\newblock
\urldef\tempurl%
\url{https://doi.org/10.1145/3434323}
\showDOI{\tempurl}


\bibitem[Hawblitzel et~al\mbox{.}(2015)]%
        {hawblitzel-ironfleet}
\bibfield{author}{\bibinfo{person}{Chris Hawblitzel}, \bibinfo{person}{Jon
  Howell}, \bibinfo{person}{Manos Kapritsos}, \bibinfo{person}{Jacob~R. Lorch},
  \bibinfo{person}{Bryan Parno}, \bibinfo{person}{Michael~L. Roberts},
  \bibinfo{person}{Srinath Setty}, {and} \bibinfo{person}{Brian Zill}.}
  \bibinfo{year}{2015}\natexlab{}.
\newblock \showarticletitle{IronFleet: Proving Practical Distributed Systems
  Correct}. In \bibinfo{booktitle}{\emph{Proceedings of the 25th Symposium on
  Operating Systems Principles}} (Monterey, California)
  \emph{(\bibinfo{series}{SOSP '15})}. \bibinfo{publisher}{Association for
  Computing Machinery}, \bibinfo{address}{New York, NY, USA},
  \bibinfo{pages}{1–17}.
\newblock
\showISBNx{9781450338349}
\urldef\tempurl%
\url{https://doi.org/10.1145/2815400.2815428}
\showDOI{\tempurl}


\bibitem[Jung et~al\mbox{.}(2018)]%
        {jung-iris}
\bibfield{author}{\bibinfo{person}{Ralf Jung}, \bibinfo{person}{Robbert
  Krebbers}, \bibinfo{person}{Jacques-Henri Jourdan}, \bibinfo{person}{Aleš
  Bizjak}, \bibinfo{person}{Lars Birkedal}, {and} \bibinfo{person}{Derek
  Dreyer}.} \bibinfo{year}{2018}\natexlab{}.
\newblock \showarticletitle{Iris from the ground up: A modular foundation for
  higher-order concurrent separation logic}.
\newblock \bibinfo{journal}{\emph{Journal of Functional Programming}}
  \bibinfo{volume}{28} (\bibinfo{year}{2018}), \bibinfo{pages}{e20}.
\newblock
\urldef\tempurl%
\url{https://doi.org/10.1017/S0956796818000151}
\showDOI{\tempurl}


\bibitem[Krogh{-}Jespersen et~al\mbox{.}(2020)]%
        {krogh-jespersen-aneris}
\bibfield{author}{\bibinfo{person}{Morten Krogh{-}Jespersen},
  \bibinfo{person}{Amin Timany}, \bibinfo{person}{Marit~Edna Ohlenbusch},
  \bibinfo{person}{Simon~Oddershede Gregersen}, {and} \bibinfo{person}{Lars
  Birkedal}.} \bibinfo{year}{2020}\natexlab{}.
\newblock \showarticletitle{Aneris: {A} Mechanised Logic for Modular Reasoning
  about Distributed Systems}. In \bibinfo{booktitle}{\emph{Programming
  Languages and Systems - 29th European Symposium on Programming, {ESOP} 2020,
  Held as Part of the European Joint Conferences on Theory and Practice of
  Software, {ETAPS} 2020, Dublin, Ireland, April 25-30, 2020, Proceedings}}.
  \bibinfo{pages}{336--365}.
\newblock
\urldef\tempurl%
\url{https://doi.org/10.1007/978-3-030-44914-8\_13}
\showDOI{\tempurl}


\bibitem[Lamport(1978)]%
        {lamport-clocks}
\bibfield{author}{\bibinfo{person}{Leslie Lamport}.}
  \bibinfo{year}{1978}\natexlab{}.
\newblock \showarticletitle{Time, Clocks, and the Ordering of Events in a
  Distributed System}.
\newblock \bibinfo{journal}{\emph{Commun. ACM}} \bibinfo{volume}{21},
  \bibinfo{number}{7} (\bibinfo{date}{July} \bibinfo{year}{1978}),
  \bibinfo{pages}{558--565}.
\newblock
\showISSN{0001-0782}
\urldef\tempurl%
\url{https://doi.org/10.1145/359545.359563}
\showDOI{\tempurl}


\bibitem[Lamport(2002)]%
        {lamport-tla}
\bibfield{author}{\bibinfo{person}{Leslie Lamport}.}
  \bibinfo{year}{2002}\natexlab{}.
\newblock \bibinfo{booktitle}{\emph{Specifying Systems: The TLA+ Language and
  Tools for Hardware and Software Engineers}}.
\newblock \bibinfo{publisher}{Addison-Wesley Longman Publishing Co., Inc.},
  \bibinfo{address}{USA}.
\newblock
\showISBNx{032114306X}


\bibitem[Leino(2010)]%
        {leino-dafny}
\bibfield{author}{\bibinfo{person}{K.~Rustan~M. Leino}.}
  \bibinfo{year}{2010}\natexlab{}.
\newblock \showarticletitle{Dafny: An Automatic Program Verifier for Functional
  Correctness}. In \bibinfo{booktitle}{\emph{Proceedings of the 16th
  International Conference on Logic for Programming, Artificial Intelligence,
  and Reasoning}} (Dakar, Senegal) \emph{(\bibinfo{series}{LPAR'10})}.
  \bibinfo{publisher}{Springer-Verlag}, \bibinfo{address}{Berlin, Heidelberg},
  \bibinfo{pages}{348–370}.
\newblock
\showISBNx{3642175104}


\bibitem[Lesani et~al\mbox{.}(2016)]%
        {lesani-chapar}
\bibfield{author}{\bibinfo{person}{Mohsen Lesani},
  \bibinfo{person}{Christian~J. Bell}, {and} \bibinfo{person}{Adam Chlipala}.}
  \bibinfo{year}{2016}\natexlab{}.
\newblock \showarticletitle{Chapar: Certified Causally Consistent Distributed
  Key-Value Stores}. In \bibinfo{booktitle}{\emph{Proceedings of the 43rd
  Annual ACM SIGPLAN-SIGACT Symposium on Principles of Programming Languages}}
  (St. Petersburg, FL, USA) \emph{(\bibinfo{series}{POPL '16})}.
  \bibinfo{publisher}{Association for Computing Machinery},
  \bibinfo{address}{New York, NY, USA}, \bibinfo{pages}{357–370}.
\newblock
\showISBNx{9781450335492}
\urldef\tempurl%
\url{https://doi.org/10.1145/2837614.2837622}
\showDOI{\tempurl}


\bibitem[Liu et~al\mbox{.}(2020)]%
        {liu-verified-rdts}
\bibfield{author}{\bibinfo{person}{Yiyun Liu}, \bibinfo{person}{James Parker},
  \bibinfo{person}{Patrick Redmond}, \bibinfo{person}{Lindsey Kuper},
  \bibinfo{person}{Michael Hicks}, {and} \bibinfo{person}{Niki Vazou}.}
  \bibinfo{year}{2020}\natexlab{}.
\newblock \showarticletitle{Verifying Replicated Data Types with Typeclass
  Refinements in Liquid Haskell}.
\newblock \bibinfo{journal}{\emph{Proc. ACM Program. Lang.}}
  \bibinfo{volume}{4}, \bibinfo{number}{OOPSLA}, Article
  \bibinfo{articleno}{216} (\bibinfo{date}{Nov.} \bibinfo{year}{2020}),
  \bibinfo{numpages}{30}~pages.
\newblock
\urldef\tempurl%
\url{https://doi.org/10.1145/3428284}
\showDOI{\tempurl}


\bibitem[Lloyd et~al\mbox{.}(2011)]%
        {lloyd-cops}
\bibfield{author}{\bibinfo{person}{Wyatt Lloyd}, \bibinfo{person}{Michael~J.
  Freedman}, \bibinfo{person}{Michael Kaminsky}, {and}
  \bibinfo{person}{David~G. Andersen}.} \bibinfo{year}{2011}\natexlab{}.
\newblock \showarticletitle{Don't Settle for Eventual: Scalable Causal
  Consistency for Wide-Area Storage with {COPS}}. In
  \bibinfo{booktitle}{\emph{Proceedings of the Twenty-Third ACM Symposium on
  Operating Systems Principles}} (Cascais, Portugal)
  \emph{(\bibinfo{series}{SOSP '11})}. \bibinfo{publisher}{Association for
  Computing Machinery}, \bibinfo{address}{New York, NY, USA},
  \bibinfo{pages}{401–416}.
\newblock
\showISBNx{9781450309776}
\urldef\tempurl%
\url{https://doi.org/10.1145/2043556.2043593}
\showDOI{\tempurl}


\bibitem[Mahajan et~al\mbox{.}(2011)]%
        {mahajan-consistency}
\bibfield{author}{\bibinfo{person}{P. Mahajan}, \bibinfo{person}{L. Alvisi},
  {and} \bibinfo{person}{M. Dahlin}.} \bibinfo{year}{2011}\natexlab{}.
\newblock \bibinfo{booktitle}{\emph{Consistency, Availability, Convergence}}.
\newblock \bibinfo{type}{{T}echnical {R}eport} TR-11-22.
  \bibinfo{institution}{Computer Science Department, University of Texas at
  Austin}.
\newblock


\bibitem[Mattern(1989)]%
        {mattern-vector-time}
\bibfield{author}{\bibinfo{person}{Friedemann Mattern}.}
  \bibinfo{year}{1989}\natexlab{}.
\newblock \showarticletitle{Virtual Time and Global States of Distributed
  Systems}. In \bibinfo{booktitle}{\emph{Parallel and Distributed Algorithms}}.
  \bibinfo{publisher}{North-Holland}, \bibinfo{pages}{215--226}.
\newblock


\bibitem[Mestanogullari et~al\mbox{.}(2015)]%
        {mestanogullari-servant}
\bibfield{author}{\bibinfo{person}{Alp Mestanogullari},
  \bibinfo{person}{S\"{o}nke Hahn}, \bibinfo{person}{Julian~K. Arni}, {and}
  \bibinfo{person}{Andres L\"{o}h}.} \bibinfo{year}{2015}\natexlab{}.
\newblock \showarticletitle{Type-Level Web APIs with Servant: An Exercise in
  Domain-Specific Generic Programming}. In
  \bibinfo{booktitle}{\emph{Proceedings of the 11th ACM SIGPLAN Workshop on
  Generic Programming}} (Vancouver, BC, Canada) \emph{(\bibinfo{series}{WGP
  2015})}. \bibinfo{publisher}{Association for Computing Machinery},
  \bibinfo{address}{New York, NY, USA}, \bibinfo{pages}{1–12}.
\newblock
\showISBNx{9781450338103}
\urldef\tempurl%
\url{https://doi.org/10.1145/2808098.2808099}
\showDOI{\tempurl}


\bibitem[Nieto et~al\mbox{.}(2022)]%
        {nieto-crdts-aneris}
\bibfield{author}{\bibinfo{person}{Abel Nieto}, \bibinfo{person}{L\'{e}on
  Gondelman}, \bibinfo{person}{Alban Reynaud}, \bibinfo{person}{Amin Timany},
  {and} \bibinfo{person}{Lars Birkedal}.} \bibinfo{year}{2022}\natexlab{}.
\newblock \showarticletitle{Modular Verification of Op-Based CRDTs in
  Separation Logic}.
\newblock \bibinfo{journal}{\emph{Proc. ACM Program. Lang.}}
  \bibinfo{volume}{6}, \bibinfo{number}{OOPSLA2}, Article
  \bibinfo{articleno}{188} (\bibinfo{date}{Oct.} \bibinfo{year}{2022}),
  \bibinfo{numpages}{29}~pages.
\newblock
\urldef\tempurl%
\url{https://doi.org/10.1145/3563351}
\showDOI{\tempurl}


\bibitem[Norell(2009)]%
        {norell-agda}
\bibfield{author}{\bibinfo{person}{Ulf Norell}.}
  \bibinfo{year}{2009}\natexlab{}.
\newblock \bibinfo{booktitle}{\emph{Dependently Typed Programming in Agda}}.
\newblock \bibinfo{publisher}{Springer Berlin Heidelberg},
  \bibinfo{address}{Berlin, Heidelberg}, \bibinfo{pages}{230--266}.
\newblock
\showISBNx{978-3-642-04652-0}
\urldef\tempurl%
\url{https://doi.org/10.1007/978-3-642-04652-0_5}
\showDOI{\tempurl}


\bibitem[Postel(1981)]%
        {RFC0793}
\bibfield{author}{\bibinfo{person}{Jon Postel}.}
  \bibinfo{year}{1981}\natexlab{}.
\newblock \bibinfo{booktitle}{\emph{Transmission Control Protocol}}.
\newblock \bibinfo{type}{STD}~7. \bibinfo{institution}{RFC Editor}.
\newblock
\showISSN{2070-1721}
\urldef\tempurl%
\url{http://www.rfc-editor.org/rfc/rfc793.txt}
\showURL{%
\tempurl}


\bibitem[Raynal et~al\mbox{.}(1991)]%
        {raynal-causal-ordering}
\bibfield{author}{\bibinfo{person}{Michel Raynal}, \bibinfo{person}{André
  Schiper}, {and} \bibinfo{person}{Sam Toueg}.}
  \bibinfo{year}{1991}\natexlab{}.
\newblock \showarticletitle{The causal ordering abstraction and a simple way to
  implement it}.
\newblock \bibinfo{journal}{\emph{Inform. Process. Lett.}}
  \bibinfo{volume}{39}, \bibinfo{number}{6} (\bibinfo{year}{1991}),
  \bibinfo{pages}{343--350}.
\newblock
\showISSN{0020-0190}
\urldef\tempurl%
\url{https://doi.org/10.1016/0020-0190(91)90008-6}
\showDOI{\tempurl}


\bibitem[{Rushby} et~al\mbox{.}(1998)]%
        {rushby-predicate-subtyping}
\bibfield{author}{\bibinfo{person}{J. {Rushby}}, \bibinfo{person}{S. {Owre}},
  {and} \bibinfo{person}{N. {Shankar}}.} \bibinfo{year}{1998}\natexlab{}.
\newblock \showarticletitle{Subtypes for specifications: predicate subtyping in
  {PVS}}.
\newblock \bibinfo{journal}{\emph{IEEE Transactions on Software Engineering}}
  \bibinfo{volume}{24}, \bibinfo{number}{9} (\bibinfo{year}{1998}),
  \bibinfo{pages}{709--720}.
\newblock
\urldef\tempurl%
\url{https://doi.org/10.1109/32.713327}
\showDOI{\tempurl}


\bibitem[Schiper et~al\mbox{.}(1989)]%
        {schiper-causal-ordering}
\bibfield{author}{\bibinfo{person}{Andr\'{e} Schiper}, \bibinfo{person}{Jorge
  Eggli}, {and} \bibinfo{person}{Alain Sandoz}.}
  \bibinfo{year}{1989}\natexlab{}.
\newblock \showarticletitle{A New Algorithm to Implement Causal Ordering}. In
  \bibinfo{booktitle}{\emph{Proceedings of the 3rd International Workshop on
  Distributed Algorithms}}. \bibinfo{publisher}{Springer-Verlag},
  \bibinfo{address}{Berlin, Heidelberg}, \bibinfo{pages}{219–232}.
\newblock
\showISBNx{3540516875}


\bibitem[{Schiper} et~al\mbox{.}(2014)]%
        {schiper-shadowdb}
\bibfield{author}{\bibinfo{person}{N. {Schiper}}, \bibinfo{person}{V. {Rahli}},
  \bibinfo{person}{R. {Van Renesse}}, \bibinfo{person}{M. {Bickford}}, {and}
  \bibinfo{person}{R.~L. {Constable}}.} \bibinfo{year}{2014}\natexlab{}.
\newblock \showarticletitle{Developing Correctly Replicated Databases Using
  Formal Tools}. In \bibinfo{booktitle}{\emph{2014 44th Annual IEEE/IFIP
  International Conference on Dependable Systems and Networks}}.
  \bibinfo{pages}{395--406}.
\newblock
\urldef\tempurl%
\url{https://doi.org/10.1109/DSN.2014.45}
\showDOI{\tempurl}


\bibitem[Schmuck(1988)]%
        {schmuck-dissertation}
\bibfield{author}{\bibinfo{person}{Frank~B Schmuck}.}
  \bibinfo{year}{1988}\natexlab{}.
\newblock \emph{\bibinfo{title}{The use of efficient broadcast protocols in
  asynchronous distributed systems}}.
\newblock \bibinfo{thesistype}{Ph.\,D. Dissertation}.
\newblock


\bibitem[Shapiro et~al\mbox{.}(2011a)]%
        {shapiro-crdts-comprehensive}
\bibfield{author}{\bibinfo{person}{Marc Shapiro}, \bibinfo{person}{Nuno
  Pregui{\c c}a}, \bibinfo{person}{Carlos Baquero}, {and}
  \bibinfo{person}{Marek Zawirski}.} \bibinfo{year}{2011}\natexlab{a}.
\newblock \bibinfo{booktitle}{\emph{{A comprehensive study of Convergent and
  Commutative Replicated Data Types}}}.
\newblock \bibinfo{type}{Research Report} RR-7506. \bibinfo{institution}{{Inria
  -- Centre Paris-Rocquencourt ; INRIA}}. \bibinfo{pages}{50} pages.
\newblock
\urldef\tempurl%
\url{https://hal.inria.fr/inria-00555588}
\showURL{%
\tempurl}


\bibitem[Shapiro et~al\mbox{.}(2011b)]%
        {shapiro-crdts}
\bibfield{author}{\bibinfo{person}{Marc Shapiro}, \bibinfo{person}{Nuno
  Pregui\c{c}a}, \bibinfo{person}{Carlos Baquero}, {and} \bibinfo{person}{Marek
  Zawirski}.} \bibinfo{year}{2011}\natexlab{b}.
\newblock \showarticletitle{Conflict-Free Replicated Data Types}. In
  \bibinfo{booktitle}{\emph{Proceedings of the 13th International Conference on
  Stabilization, Safety, and Security of Distributed Systems}} (Grenoble,
  France) \emph{(\bibinfo{series}{SSS'11})}.
  \bibinfo{publisher}{Springer-Verlag}, \bibinfo{address}{Berlin, Heidelberg},
  \bibinfo{pages}{386–400}.
\newblock
\showISBNx{9783642245497}


\bibitem[van Renesse(1993)]%
        {van-renesse-controversy}
\bibfield{author}{\bibinfo{person}{Robbert van Renesse}.}
  \bibinfo{year}{1993}\natexlab{}.
\newblock \showarticletitle{Causal Controversy at Le Mont St.-Michel}.
\newblock \bibinfo{journal}{\emph{SIGOPS Oper. Syst. Rev.}}
  \bibinfo{volume}{27}, \bibinfo{number}{2} (\bibinfo{date}{April}
  \bibinfo{year}{1993}), \bibinfo{pages}{44–53}.
\newblock
\showISSN{0163-5980}
\urldef\tempurl%
\url{https://doi.org/10.1145/155848.155857}
\showDOI{\tempurl}


\bibitem[Vazou et~al\mbox{.}(2018)]%
        {vazou-theorem-proving-for-all}
\bibfield{author}{\bibinfo{person}{Niki Vazou}, \bibinfo{person}{Joachim
  Breitner}, \bibinfo{person}{Rose Kunkel}, \bibinfo{person}{David Van~Horn},
  {and} \bibinfo{person}{Graham Hutton}.} \bibinfo{year}{2018}\natexlab{}.
\newblock \showarticletitle{Theorem Proving for All: Equational Reasoning in
  Liquid Haskell (Functional Pearl)}. In \bibinfo{booktitle}{\emph{Proceedings
  of the 11th ACM SIGPLAN International Symposium on Haskell}} (St. Louis, MO,
  USA) \emph{(\bibinfo{series}{Haskell 2018})}. \bibinfo{publisher}{Association
  for Computing Machinery}, \bibinfo{address}{New York, NY, USA},
  \bibinfo{pages}{132–144}.
\newblock
\showISBNx{9781450358354}
\urldef\tempurl%
\url{https://doi.org/10.1145/3242744.3242756}
\showDOI{\tempurl}


\bibitem[Vazou et~al\mbox{.}(2013)]%
        {vazou-abstract-refinements}
\bibfield{author}{\bibinfo{person}{Niki Vazou}, \bibinfo{person}{Patrick~Maxim
  Rondon}, {and} \bibinfo{person}{Ranjit Jhala}.}
  \bibinfo{year}{2013}\natexlab{}.
\newblock \showarticletitle{Abstract Refinement Types}. In
  \bibinfo{booktitle}{\emph{Programming Languages and Systems - 22nd European
  Symposium on Programming, {ESOP} 2013, Held as Part of the European Joint
  Conferences on Theory and Practice of Software, {ETAPS} 2013, Rome, Italy,
  March 16-24, 2013. Proceedings}}. \bibinfo{pages}{209--228}.
\newblock
\urldef\tempurl%
\url{https://doi.org/10.1007/978-3-642-37036-6\_13}
\showDOI{\tempurl}


\bibitem[Vazou et~al\mbox{.}(2014)]%
        {vazou-lh}
\bibfield{author}{\bibinfo{person}{Niki Vazou}, \bibinfo{person}{Eric~L.
  Seidel}, \bibinfo{person}{Ranjit Jhala}, \bibinfo{person}{Dimitrios
  Vytiniotis}, {and} \bibinfo{person}{Simon Peyton-Jones}.}
  \bibinfo{year}{2014}\natexlab{}.
\newblock \showarticletitle{Refinement Types for Haskell}. In
  \bibinfo{booktitle}{\emph{Proceedings of the 19th ACM SIGPLAN International
  Conference on Functional Programming}} (Gothenburg, Sweden)
  \emph{(\bibinfo{series}{ICFP '14})}. \bibinfo{publisher}{Association for
  Computing Machinery}, \bibinfo{address}{New York, NY, USA},
  \bibinfo{pages}{269–282}.
\newblock
\showISBNx{9781450328739}
\urldef\tempurl%
\url{https://doi.org/10.1145/2628136.2628161}
\showDOI{\tempurl}


\bibitem[Vazou et~al\mbox{.}(2017)]%
        {vazou-refinement-reflection}
\bibfield{author}{\bibinfo{person}{Niki Vazou}, \bibinfo{person}{Anish
  Tondwalkar}, \bibinfo{person}{Vikraman Choudhury}, \bibinfo{person}{Ryan~G.
  Scott}, \bibinfo{person}{Ryan~R. Newton}, \bibinfo{person}{Philip Wadler},
  {and} \bibinfo{person}{Ranjit Jhala}.} \bibinfo{year}{2017}\natexlab{}.
\newblock \showarticletitle{Refinement Reflection: Complete Verification with
  SMT}.
\newblock \bibinfo{journal}{\emph{Proc. ACM Program. Lang.}}
  \bibinfo{volume}{2}, \bibinfo{number}{POPL}, Article \bibinfo{articleno}{53}
  (\bibinfo{date}{Dec.} \bibinfo{year}{2017}), \bibinfo{numpages}{31}~pages.
\newblock
\urldef\tempurl%
\url{https://doi.org/10.1145/3158141}
\showDOI{\tempurl}


\bibitem[Wenzel et~al\mbox{.}(2008)]%
        {wenzel-isabelle}
\bibfield{author}{\bibinfo{person}{Makarius Wenzel},
  \bibinfo{person}{Lawrence~C. Paulson}, {and} \bibinfo{person}{Tobias
  Nipkow}.} \bibinfo{year}{2008}\natexlab{}.
\newblock \showarticletitle{The Isabelle Framework}. In
  \bibinfo{booktitle}{\emph{Theorem Proving in Higher Order Logics}},
  \bibfield{editor}{\bibinfo{person}{Otmane~Ait Mohamed},
  \bibinfo{person}{C{\'e}sar Mu{\~{n}}oz}, {and} \bibinfo{person}{Sofi{\`e}ne
  Tahar}} (Eds.). \bibinfo{publisher}{Springer Berlin Heidelberg},
  \bibinfo{address}{Berlin, Heidelberg}, \bibinfo{pages}{33--38}.
\newblock
\showISBNx{978-3-540-71067-7}


\bibitem[Wilcox et~al\mbox{.}(2015)]%
        {wilcox-verdi}
\bibfield{author}{\bibinfo{person}{James~R. Wilcox}, \bibinfo{person}{Doug
  Woos}, \bibinfo{person}{Pavel Panchekha}, \bibinfo{person}{Zachary Tatlock},
  \bibinfo{person}{Xi Wang}, \bibinfo{person}{Michael~D. Ernst}, {and}
  \bibinfo{person}{Thomas Anderson}.} \bibinfo{year}{2015}\natexlab{}.
\newblock \showarticletitle{Verdi: A Framework for Implementing and Formally
  Verifying Distributed Systems}. In \bibinfo{booktitle}{\emph{Proceedings of
  the 36th ACM SIGPLAN Conference on Programming Language Design and
  Implementation}} (Portland, OR, USA) \emph{(\bibinfo{series}{PLDI '15})}.
  \bibinfo{publisher}{Association for Computing Machinery},
  \bibinfo{address}{New York, NY, USA}, \bibinfo{pages}{357–368}.
\newblock
\showISBNx{9781450334686}
\urldef\tempurl%
\url{https://doi.org/10.1145/2737924.2737958}
\showDOI{\tempurl}


\bibitem[Xi and Pfenning(1998)]%
        {xi-array-dependent}
\bibfield{author}{\bibinfo{person}{Hongwei Xi} {and} \bibinfo{person}{Frank
  Pfenning}.} \bibinfo{year}{1998}\natexlab{}.
\newblock \showarticletitle{Eliminating Array Bound Checking through Dependent
  Types}. In \bibinfo{booktitle}{\emph{Proceedings of the ACM SIGPLAN 1998
  Conference on Programming Language Design and Implementation}} (Montreal,
  Quebec, Canada) \emph{(\bibinfo{series}{PLDI '98})}.
  \bibinfo{publisher}{Association for Computing Machinery},
  \bibinfo{address}{New York, NY, USA}, \bibinfo{pages}{249–257}.
\newblock
\showISBNx{0897919874}
\urldef\tempurl%
\url{https://doi.org/10.1145/277650.277732}
\showDOI{\tempurl}


\end{thebibliography}

\end{document}